\documentclass[lettersize,journal]{IEEEtran}
\usepackage{amsmath,amsfonts}  
\usepackage{algorithmic}
\usepackage{algorithm}
\usepackage{array}
\usepackage{textcomp}
\usepackage{stfloats}
\usepackage{url}
\usepackage{verbatim}
\usepackage{graphicx}
\usepackage{cite}
\usepackage{lipsum}
\usepackage{subfigure}
\usepackage{hyperref} 
\usepackage{float}
\usepackage{calc}
\usepackage{booktabs}

\usepackage[acronym,nomain]{glossaries}
\renewcommand{\glossarysection}[2][]{}

\newglossarystyle{mystyle}
{%
  {% start of glossary
    \setlength{\maxlength}{0pt}%
    \forglsentries[\currentglossary]{\thislabel}%
    {%
       \settowidth{\thislength}{\glsentryshort{\thislabel}}%
       \ifdim\thislength>\maxlength
         \setlength{\maxlength}{\thislength}%
       \fi
    }%
    \settowidth{\thislength}{\hspace{1.5em}\hspace{1em}}%
    \setlength{\glsdescwidth}{\linewidth-\maxlength-\thislength-2\tabcolsep}%
    \begin{tabular}{l@{\hspace{1.5em}\hspace{1em}}p{\glsdescwidth}}
    \toprule
    \multicolumn{1}{l}{\textbf{Abbreviation}} &
    \multicolumn{1}{@{}l}{\textbf{Description}}\\%
    \midrule
  }%
  {% end of glossary
     \bottomrule
     \end{tabular}%
  }%
  \renewcommand*{\glsgroupheading}[1]{}%
}

\makeglossaries
\newacronym{3gpp}{3GPP}{3rd generation partnership project}
\newacronym{6g}{6G}{6th generation}
\newacronym{5g}{5G}{5th generation}
\newacronym{3g}{3G}{third generation} 
\newacronym{4g}{4G}{fourth generation}

\newacronym{ack}{ACK}{acknowledgment}
\newacronym{acm}{ACM}{adaptive coding and modulation}
\newacronym{arq}{ARQ}{automatic repeat request}
\newacronym{ap}{AP}{access point}
\newacronym{ai}{AI}{artificial intelligence}
\newacronym{af}{AF}{amplify-and-forward}
\newacronym{aoi}{AoI}{age of information}
\newacronym{ac}{AC}{access category}
\newacronym{amd}{AMD}{acknowledged mode data}
\newacronym{am}{AM}{acknowledged mode}
\newacronym{ami}{AMI}{accumulated mutual information}
\newacronym{awgn}{AWGN}{additive white Gaussian noise}
\newacronym{asr}{ASR}{achievable sum rate}

\newacronym{b5g}{B5G}{beyond 5G}
\newacronym{ber}{BER}{bit error rate}
\newacronym{bler}{BLER}{bLock error rate}
\newacronym{bpsk}{BPSK}{binary phase shift keying}
\newacronym{btfdps}{BTFDPS}{block time-frequency domain packet scheduling}
\newacronym{bpu}{BPU}{baseband processing unit}
\newacronym{blc}{BLC}{bit level combining}
\newacronym{back}{BAck}{block \gls{ack}}
\newacronym{bs}{BS}{base station}
\newacronym{bch}{BCH}{Bose–Chaudhuri–Hocquenghem}
\newacronym{bps}{bps}{bits per second}

\newacronym{cqi}{CQI}{channel quality indicator}
\newacronym{crc}{CRC}{cyclic redundancy check}
\newacronym{csi}{CSI}{channel state information}
\newacronym{cr}{CR}{cognitive radio}
\newacronym{cfr}{CFR}{channel frequency response}
\newacronym{cpc}{CPC}{complementary punctured codes}
\newacronym{cran}{CRAN}{cloud-radio access network}
\newacronym{cbg}{CBG}{codeblock-group}
\newacronym{cdma}{CDMA}{code division multiple access}
\newacronym{cc}{CC}{Chase combining}
\newacronym{cd}{CD}{code domain}
\newacronym{cnoma}{C-NOMA}{cooperative \gls{noma}}
\newacronym{crsic}{C-RSIC}{cooperative retroactive \gls{sic}}
\newacronym{crstlc}{CR-STLC}{constellation-rotated spacetime line code}

\newacronym{d2d}{D2D}{device-to-device}
\newacronym{di}{DI}{deferred iterations}
\newacronym{dl}{DL}{downlink}
\newacronym{dcf}{DCF}{decode-and-forward}
\newacronym{df}{DF}{demodulated-and-forward}
\newacronym{dfrc}{DFRC}{dual functional radar communication}
\newacronym{dd}{DD}{delay-Doppler}

\newacronym{es}{ES}{early stop}
\newacronym{ed}{ED}{error detection}
\newacronym{eharq}{E-HARQ}{early HARQ}
\newacronym{enb}{eNB}{evolved node B}
\newacronym{e-utra}{E-UTRA}{evolved universal terrestrial radio access}
\newacronym{eutran}{EUTRAN}{evolved universal terrestrial radio access network}
\newacronym{ee}{EE}{energy efficiency}

\newacronym{fec}{FEC}{forward error correction}
\newacronym{fifo}{FIFO}{first in, first out}
\newacronym{fdd}{FDD}{frequency division duplexing}
\newacronym{fps}{fps}{frames per second}
\newacronym{fer}{FER}{frame error rate}
\newacronym{fm}{$F_M$}{feedback message}
\newacronym{fu}{FU}{far-user}
\newacronym{fdma}{FDMA}{frequency division multiple access}

\newacronym{gop}{GoP}{group of pictures}
\newacronym{gnom}{GNOM}{generalized \gls{nom}}
\newacronym{gsm}{GSM}{global system for mobile communications}
\newacronym{gbn}{GBN}{go-back-N}

\newacronym{harq}{HARQ}{hybrid-ARQ}
\newacronym{hevc}{HEVC}{high efficiency video coding}
\newacronym{hiho}{HIHO}{hard-input hard-output}
\newacronym{hr}{HR}{high-rate}
\newacronym{hd}{HD}{half duplex}
\newacronym{hsdpa}{HSDPA}{high speed downlink packet access}
\newacronym{hsupa}{HSUPA}{high speed uplink packet access}
\newacronym{hspa+}{HSPA+}{evolved high speed packet access}
\newacronym{hdd}{HDD}{hard decision decoding}

\newacronym{iot}{IoT}{Internet of Things}
\newacronym{itu}{ITU}{international telecommunication union}
\newacronym{ici}{ICI}{inter-carrier interference}
\newacronym{ipic}{IPIC}{inter-packet interference cancellation}
\newacronym{ic}{IC}{interference cancellation}
\newacronym{its}{ITS}{intelligent transportation system}
\newacronym{ir}{IR}{incremental redundancy}
\newacronym{iid}{i.i.d.}{independent and identically distributed}
\newacronym{irs}{IRS}{intelligent reflecting surface}
\newacronym{imt}{IMT}{international mobile telecommunications}
\newacronym{isac}{ISaC}{integrated sensing and communication}

\newacronym{jml}{JML}{joint maximum likelihood}

\newacronym{ldpc}{LDPC}{low density parity check}
\newacronym{lifo}{LIFO}{last in, first out}
\newacronym{lte}{LTE}{long term evolution}
\newacronym{leos}{LEOS}{low-earth orbit satellite}
\newacronym{lr}{LR}{low-rate}
\newacronym{lfsr}{LFSR}{linear  feedback  shift register}
\newacronym{llr}{LLR}{log-likelihood ratio}
\newacronym{lrsic}{L-RSIC}{local retroactive \gls{sic}}
\newacronym{lsb}{LSB}{least significant bit}

\newacronym{msg3}{MSG3}{message-3}
\newacronym{mimo}{MIMO}{multiple-input multiple-output}
\newacronym{mmimo}{mMIMO}{massive \gls{mimo}}
\newacronym{mmwave}{mm-Wave}{millimeter wave}
\newacronym{mac}{MAC}{medium access control}
\newacronym{mib}{MIB}{management information base}
\newacronym{maf}{MAF}{moving average filter}
\newacronym{mpeg}{MPEG}{moving picture experts group}
\newacronym{mse}{MSE}{mean squared error}
\newacronym{mld}{MLD}{maximum likelihood detection}
\newacronym{mbs}{MBS}{macro base station}
\newacronym{mpdu}{MPDU}{\gls{mac} protocol data unit}
\newacronym{mm}{mm}{millimeter}
\newacronym{mrc}{MRC}{maximum ratio combining}
\newacronym{mpsk}{MPSK}{M-ary phase shift keying}
\newacronym{mcs}{MCS}{modulation and coding scheme}
\newacronym{mtc}{MTC}{machine type communication}
\newacronym{mmtc}{mMTC}{massive machine type communication}
\newacronym{md}{MD}{minimum distance}
\newacronym{ml}{ML}{maximum likelihood}
\newacronym{msb}{MSB}{most significant bit}

\newacronym{nb}{NB}{narrow band}
\newacronym{nack}{NACK}{negative acknowledgment}
\newacronym{noma}{NOMA}{non-orthogonal multiple access}
\newacronym{nom}{NOM}{non-orthogonal multiplexing}
\newacronym{nr}{NR}{new radio}
\newacronym{nu}{NU}{near-user}

\newacronym{ofdm}{OFDM}{orthogonal frequency-division multiplexing}
\newacronym{oma}{OMA}{orthogonal multiple access}
\newacronym{otfs}{OTFS}{orthogonal time–frequency space}

\newacronym{per}{PER}{packet error rate}
\newacronym{ptd}{PTD}{packet time delay}
\newacronym{pdr}{PDR}{packet drop rate}
\newacronym{prr}{PRR}{packets repair and recovery}
\newacronym{pr}{PR}{packet recovery}
\newacronym{pec}{PEC}{parity error checking}
\newacronym{pi}{PI}{percentage improvement}
\newacronym{phy}{PHY}{physical layer}
\newacronym{prb}{PRB}{physical resource block}
\newacronym{pfa}{PFA}{probability of false alarm}
\newacronym{pm}{PM}{probability of miss}
\newacronym{psnr}{PSNR}{peak signal to noise ratio}
\newacronym{psk}{PSK}{phase shift keying}
\newacronym{pep}{PEP}{packet error probability}
\newacronym{pd}{PD}{power domain}
\newacronym{psdu}{PSDU}{physical data service unit}
\newacronym{pdu}{PDU}{protocol data unit}
\newacronym{pps}{pps}{packets per second}
\newacronym{pdf}{PDF}{probability density function}

\newacronym{qam}{QAM}{quadrature amplitude modulation}
\newacronym{qos}{QoS}{quality of service}
\newacronym{qp}{QP}{quantization parameter}
\newacronym{qpsk}{QPSK}{quadrature phase shift keying}

\newacronym{rlc}{RLC}{radio link control}
\newacronym{rsic}{RSIC}{retroactive \gls{sic}}
\newacronym{rf}{RF}{radio frequency}
\newacronym{rrh}{RRH}{remote  radio  head}
\newacronym{rau}{RAU}{radio access unit}
\newacronym{rl}{RL}{reinforcement learning}
\newacronym{rrc}{RRC}{radio resources control}
\newacronym{rtt}{RTT}{round trip time}
\newacronym{ran}{RAN}{radio access network}
\newacronym{rsma}{RSMA}{rate-splitting multiple access}
\newacronym{rcs}{RCS}{radar cross section}

\newacronym{ser}{SER}{symbol error rate}
\newacronym{sic}{SIC}{successive interference cancellation}
\newacronym{snr}{SNR}{signal to noise ratio}
\newacronym{sr-harq}{SR-HARQ}{selective repeat \gls{harq}}
\newacronym{sc}{SC}{superposition coding}
\newacronym{swipt}{SWIPT}{simultaneous wireless information and power transmission}
\newacronym{sbs}{SBS}{small base station}
\newacronym{slc}{SLC}{symbol level combining}
\newacronym{snmp}{SNMP}{simple network management protocol}
\newacronym{sta}{STA}{wireless station}
\newacronym{ssrc}{SSRC}{station short retry count}
\newacronym{slrc}{SLRC}{station long retry count}
\newacronym{spid}{SPID}{subpacket identifier}
\newacronym{sn}{SN}{sequence number}
\newacronym{scdma}{SCDMA}{synchronous \gls{cdma}}
\newacronym{sinr}{SINR}{signal to interference plus noise ratio}
\newacronym{sdd}{SDD}{soft decision decoding}
\newacronym{sr}{SR}{selective repeat}
\newacronym{sw}{SW}{stop-and-wait}
\newacronym{std}{STD}{standard deviation}
\newacronym{siso}{SISO}{soft-input soft-output}
\newacronym{sd}{SD}{spatial decoding}
\newacronym{spc}{SPC}{short packet communication}

\newacronym{tpc}{TPC}{turbo product code}
\newacronym{tcp}{TCP}{transmission control protocol}
\newacronym{tc}{TC}{traffic category}
\newacronym{tb}{TB}{transport block}
\newacronym{td}{TD}{temporal decoding}  
\newacronym{tti}{TTI}{transmission time interval}
\newacronym{tdps}{TDPS}{time domain packet scheduling}
\newacronym{tdd}{TDD}{time division duplexing}
\newacronym{tcc}{TCC}{turbo convolutional code}
\newacronym{tnr}{TNR}{transmission number relaying}
\newacronym{tm}{TM}{transmissions per message}
\newacronym{tn}{TN}{terrestrial network}
\newacronym{ts}{TS}{transmission slot}
\newacronym{tdma}{TDMA}{time division multiple access}

\newacronym{ul}{UL}{uplink}
\newacronym{ue}{UE}{user equipment}
\newacronym{umts}{UMTS}{universal mobile telecommunications system}
\newacronym{urllc}{URLLC}{ultra reliable low latency communications}
\newacronym{um}{UM}{unacknowledged mode}
\newacronym{uav}{UAV}{unmanned aerial vehicle}

\newacronym{voip}{VoIP}{voice  over  internet  protocol} 
\newacronym{v2v}{V2V}{vehicle to vehicle}
\newacronym{v2i}{V2I}{vehicle to infrastructure}
\newacronym{vlc}{VLC}{visible light communication}

\newacronym{wifi}{WiFi}{wireless fidelity}
\newacronym{wran}{WRAN}{wireless regional area network}
\newacronym{wan}{WAN}{wide area network}
\newacronym{wpan}{WPAN}{wireless personal area network}
\newacronym{wlan}{WLAN}{wireless local area network}
\newacronym{wimax}{WiMAX}{worldwide  interoperability  for  microwave access}
\newacronym{wcdma}{W-CDMA}{wideband-CDMA}
\newacronym{wap}{WAP}{wireless access point}
\newacronym{wsn}{WSN}{wireless sensor network}

\newacronym{ntn}{NTN}{non-terrestrial network}
\newacronym{los}{LOS}{line of sight}

\newacronym{zf}{ZF}{zero-forcing}

\newacronym{wpans}{WPANs}{wireless personal area networks}

\newacronym{tpcs}{TPCs}{turbo product codes}
\newacronym{lfsrs}{LFSRs}{linear  feedback  shift registers}
\newacronym{bsn}{BSN}{block sequence number}
\newlength\maxlength
\newlength\thislength

\hypersetup{
    colorlinks=true,
    linkcolor=black, % Change the color of abbreviation
    filecolor=magenta,      
    urlcolor=cyan,
    pdftitle={Overleaf Example},
    pdfpagemode=FullScreen,
    }

\begin{document}

\title{Performance Analysis of Pair-wise Symbol Detection in Uplink NOMA-ISaC Systems}

\author{
Haofeng~Liu,~\IEEEmembership{Graduate~Student~Member,~IEEE,}~Emad~Alsusa,~\IEEEmembership{Senior~Member,~IEEE,}
and~Arafat~Al-Dweik,~\IEEEmembership{Senior~Member,~IEEE}% <-this % stops a space

\thanks{Haofeng Liu and Emad Alsusa are with the Department of Electrical and Electronic Engineering, The University of Manchester, Manchester, M13 9PL, United Kingdom. (email: haofeng.liu@postgrad.manchester.ac.uk, e.alsusa@manchester.ac.uk). 

Arafat Al-Dweik is with 6G Reserch Center, Department of Computer and Communications Engineering, Khalifa University, Abu Dhabi, 127788, United Arab Emirates (e-mail: arafat.dweik@ku.ac.ae, dweik@fulbrightmail.org)}
}

% \markboth{Journal of XXX,~Vol.~XXX, No.~XX, XXX~20XX}%

\maketitle

\begin{abstract}

This paper investigates the bit error rate (BER) and outage probability performance of integrated sensing and communication (ISaC) in uplink non-orthogonal multiple access (NOMA) based Internet of Things (IoT) systems. Specifically, we consider an ISaC system where the radar signal is designed to be orthogonal to the communication signal over two symbol periods so that its interference on the communication signal is completely eliminated when detecting the data in pairs of consecutive symbols. This is akin to multi-symbol rate NOMA systems except in this case as the radar bears no data, its waveform is manipulated to be orthogonal to the transmitted communication signal. To eliminate potential decision ambiguity during the pair-wise data detection, a constant phase-offset between adjacent communication symbols is applied at the transmitter. The performance of such a system is analyzed through deriving analytical expressions for the exact BER of zero-forcing (ZF) based receivers. In addition, close-form expressions for the upper BER bound and the outage probability for both ZF and the joint maximum likelihood (JML) receivers are presented. The results show that the derived expressions are perfectly matched with the simulation results. The obtained expressions provide an insight into the performance of this novel ISaC system including demonstrating the impact of various parameters and showing how the ZF receiver provides a useful trade-off between performance and complexity relative to the JML receiver.

\end{abstract}

\begin{IEEEkeywords}
Uplink non-orthogonal multiple access (NOMA), integrated sensing and communication (ISaC), bit error rate (BER), outage probability.
\end{IEEEkeywords}

\section{Introduction}
\label{sec:s1}
\IEEEPARstart{T}{he} advent of \gls{6g} communication systems brings a paradigm shift toward integrating communication with radar detection capabilities, which is crucial for applications such as the location of \gls{uav} and automated industrial processes \cite{Intro_1}. In addition, \gls{iot} devices play a key role in perfecting communication and radar detection technologies with their ubiquitous sensors and interconnections. By leveraging a dense network of \gls{iot} sensors, \gls{6g} systems can achieve unprecedented levels of environmental awareness and data collection. \cite{Intro_1_1} However, the growing number of mobile devices presents a significant challenge in satisfying both the demands of communication and radar systems within the limited spectral resources. Traditional orthogonal access methods such as \gls{tdma} and \gls{fdma}, which split the spectrum into discrete time slots or frequency bands, are unable to achieve the required levels of efficiency. Hence, there is a growing trend in using the same time-frequency resources for simultaneous sensing and communication, an innovative approach more often referred to as \gls{isac} or \gls{dfrc} \cite{Intro_2}, which has been shown to offer promising enhancements in spectrum use.

\gls{isac} represents a significant technological convergence in cooperating radar technology with communication systems into a unified platform. This facilitates data collection, transmission, and processing in radar and communication networks \cite{Intro_4}. \gls{isac} can have many advantages, including increased spectrum efficiency, enhanced functionality, and improved reliability, as well as many potential applications such as the \gls{iot}, smart cities, industrial automation, and healthcare. The unique advantages and practical benefits of \gls{isac} underscore its critical role in advancing the spectral efficiency and capabilities of modern technologies \cite{Intro_5}.

Similarly, \gls{noma} is a potential technology for \gls{6g} communication networks \cite{Intro_6}. This approach is distinct from traditional approaches in that it uses the principle of superposition to multiplex users at the same time and frequency resource, hence improving spectral utilization and efficiency. In doing so, \gls{noma} can significantly increase the system capacity without requiring additional spectral resources. Such advantage of \gls{noma} makes it a key technology for the future \gls{6g} networks, which is expected to meet the growing demand for a rapid increase in mobile devices and increased data rate requirements \cite{Intro_8}.

\subsection{Related work}
Many of the existing studies on \gls{noma} systems focus on the \gls{ber} performance \cite{w1,w2,w3,w4}. For example, in \cite{w1}, the authors evaluate the \gls{ber} performance of a \gls{noma} system with two users having equal distances from the base station under various channel conditions including \gls{awgn} and fading environments such as Rayleigh, Rician, Nakagami-$m$ and Weibull. On the other hand, the authors in \cite{w2} investigated the performance of the \gls{ser} and the \gls{ber} in \gls{noma} systems considering the direction of transmission, the number of information streams, and the effects of channel coding. Furthermore, a joint Gray coding scheme was proposed to significantly reduce \gls{ser} and \gls{ber} in downlink transmissions. In \cite{w3}, the authors focused on enhancing \gls{noma}, and presented exact and asymptotic \gls{ber} expressions under Rayleigh fading conditions for \gls{noma} systems with varying user counts, receiving antennas, and modulation types, including \gls{bpsk} and \gls{qam}. Additionally, the study derived bounds for the power coefficients to ensure fairness and address constellation ambiguity for systems with two and three users across different modulation orders. The findings highlighted that the range of feasible power coefficients narrows as either the number of users or the modulation complexity increases, exacerbating inter-user interference and impacting \gls{ber}. The authors in \cite{w4} addressed the gaps in \gls{noma} \gls{ber} analysis by considering the transformation of the noise \gls{pdf} after using \gls{sic}. The first derivation of the exact noise \gls{pdf} post-\gls{sic} as a truncated Gaussian mixture was presented, and highlighted the need for different models for successful and unsuccessful \gls{sic} scenarios which significantly improved the accuracy of \gls{ber} calculation in \gls{noma} systems.

The combination of \gls{noma} and \gls{isac} has also been studied in \cite{w5,w6,w7,w8}. The authors in \cite{w5} introduced a \gls{noma}-assisted \gls{otfs} network for \gls{isac}, utilizing \gls{uav} as air \gls{bs}. The system leveraged \gls{isac} to gather positional and velocity data from the user's echo signals, enhancing system performance through non-orthogonal power allocation tailored for optimal \gls{asr}. Furthermore, the paper developed a robust power allocation strategy that addresses max-min fairness and maximum \gls{asr} issues under both perfect and imperfect \gls{csi}. The simulation results affirmed the superiority of the proposed \gls{noma}-assisted \gls{otfs}-\gls{isac} system in achieving higher rates compared to other systems. A novel semi-\gls{isac} framework was proposed in \cite{w6} that offered a flexible bandwidth allocation between exclusive wireless communication, radar detection and integrated \gls{isac} transmission. The paper compared the bandwidth efficiency of semi-\gls{isac} networks that transition from \gls{oma} to \gls{noma} by assessing the probability of outage, the ergodic communication rates, and the ergodic radar estimation information rate. Simulation results showed that \gls{noma}-based Semi-\gls{isac} provides distinct diversity orders for near and far users. In \cite{w7}, the authors studied a \gls{noma}-based \gls{isac} system in which the base station simultaneously transmits superimposed communication and sensing signals. They proposed an interference cancelation scheme that optimizes these signals to achieve the desired sensing pattern while meeting the \gls{sic} and communication user rate requirements. The simulation results showed that the proposed system outperforms the conventional \gls{isac}. In paper \cite{w8}, the integration of \gls{noma} with \gls{iot} networks was explored to enhance global coverage and resource efficiency in satellite communications. The study introduced a dual approach to secure precoding in a low earth orbit satellite system, addressing both perfect and imperfect \gls{csi} of potential eavesdroppers. For each scenario, a joint precoding optimization problem was formulated to maximize the sum secrecy rate using strategies such as artificial jamming. The simulation results demonstrated that the proposed \gls{noma}-\gls{isac} scheme outperforms traditional \gls{tdma}in terms of sum secrecy rate, while still maintaining effective sensing capabilities.

On the uplink, several \gls{noma}-based \gls{isac} systems were studied \cite{w9,w10,w11}. The authors in \cite{w9} derived novel expressions for the probability of communication and sensing rates. The simulation results showed that the \gls{isac} systems outperform traditional frequency division sensing and communications in sensing rates without compromising the communication quality. The simulation results highlighted the effectiveness of \gls{noma} in \gls{isac} performance enhancement. In \cite{w10}, the authors provided exact and asymptotic analyzes of the outage probabilities for uplink \gls{ue} and derived the probability of successful detection for the sensing performance. The results indicated that \gls{isac} \gls{noma} outperforms traditional \gls{isac} systems in both communication reliability and sensing accuracy. The work in \cite{w11} focused on the uplink signal and communication performance of \gls{noma}-based \gls{isac} systems, particularly examining the impact of \gls{sic} ordering on system performance. By deriving the diversity orders, high \gls{snr} slopes, and high-\gls{snr} power offsets, the simulation results showed that the \gls{sic} order significantly affects the system's spectral efficiency and communication reliability through modifications in high-\gls{snr} power offsets and array gains.

Several studies have shown that rotating the constellation diagrams or applying different phase shifts for different \glspl{ue} are promising methods to improve the performance of the \gls{noma} systems \cite{w12,w13,w14,w15,w16}. In \cite{w12}, the authors derived semi-analytical expressions of \gls{ser} and \gls{ber} for the \gls{jml} detection of two \gls{qpsk} signals with varying phase offsets between the users. The closed-form union bound expressions for the user \gls{ser} and \gls{ber} across any phase offset and power ratio are also presented, highlighting the unique error rate dynamics that the detection of \gls{jml} exhibits per user. In \cite{w13} a \gls{noma} downlink system is considered over block fading channels. The authors proposed new schemes that employ $n$-dimensional constellations from the same algebraic lattices in a number field, allowing each user to achieve a full diversity gain without the need for \gls{sic}. The paper analyzes and derives upper bounds for the minimum product distances within these schemes under any power allocation factor. The simulation results showed an improved \gls{ser} compared to traditional \gls{noma} approaches. In \cite{w14}, the research focused on the \gls{noma} uplink that uses \gls{crstlc} to communicate with a dual antenna access point. The paper presented a mathematical analysis of the \gls{ber} performance and spatial diversity order in a two-user uplink \gls{crstlc} \gls{noma} setting. Furthermore, the study compared the dynamic and fixed rotation angle optimization techniques, showing that fixed rotation performs similarly to dynamic rotation in practical \gls{snr} scenarios with lower computational cost. The authors in \cite{w15} presented a novel investigation of optimal inter-constellation rotation for the uplink \gls{noma}, focusing on improving signal distinctiveness using the \gls{md} criterion. This study derived closed-form expressions for the \gls{md}-maximizing rotation angle and its consequently largest \gls{md} between joint constellation points across the common modulation schemes. In \cite{w16}, the authors introduced a novel approach to improve the uplink performance of \gls{noma} by implementing constellation rotation, particularly by improving the efficiency of \gls{sic} receivers. An optimal closed-form solution is achieved to maximize the entropy through variational approximation. Theoretical analysis and simulation results reveal that this scheme surpasses traditional \gls{noma} in capacity and bit error rate, especially when the receiver has fewer antennas or when there is a minimal channel gain difference between the users.

\subsection{Motivation and Contribution}

From the surveyed literature, the theoretical analysis of \gls{ber} and the probability of outage has been carried out in a wide range of scenarios. However, the aforementioned papers all considered detection on a symbol by symbol basis. On the contrary, we proposed a novel signaling approach in which the receiver detects the signals over two periods of symbols \cite{my} to enhance the effectiveness of canceling mutual interference between the radar and the communication system. This is achieved by designing the radar waveform to be orthogonal to the communication signals over two symbol periods to make it possible to eliminate interference between the two subsystems. We have showed that the proposed ISaC system offers superior performance for both communication and radar systems. Nevertheless, the results provided in our previous work were simulation based only. Therefore, in this study, our aim is to present an analytical framework of our proposed \gls{isac} system to provide deeper insights into this system. Specifically, we demonstrate an example of the decision bounds and regions, then derive semi-analytical expressions for the \gls{ber} of the \glspl{ue} with a fixed phase shift in alternate symbol periods. In addition, closed-form expressions for the \gls{ber} upper bound and the outage probability are presented. The detailed contributions can be summarized as follows:

\begin{enumerate}

\item We present a mathematical framework for our \gls{isac} signaling approach and explain the principle of pair-wise symbol detection.

\item We demonstrate the decision regions with fixed rotated phase shifts and derive semi-analytical \gls{ber} expressions for the \gls{zf} receiver.

\item We derive close-form upper bound \gls{ber} expressions for the \gls{jml} and \gls{zf} receivers with fixed rotated phase shift.

\item We derive closed-form expressions for the capacity outage probability of the \gls{jml} and \gls{zf} receivers, respectively, and average the \gls{jml} receiver over the Gamma distribution and over the inverse-Gamma distribution for the \gls{zf} receiver.
\item We present comprehensive results and discussion of the \gls{isac} system under consideration and discuss various performance trade-offs.
\end{enumerate}

\textit{Notation}: In this paper, the following notation is used. Lowercase bold letters denote vectors (e.g., $\mathbf{h}$), and uppercase bold letters denote matrix (e.g., $\mathbf{H}$). For a vector $\mathbf{a}$, $\mathbf{a}^{\dag}$, $\mathbf{a}^*$, $\mathbf{a}^H$, and $\left \| \mathbf{a} \right \|$ represent the transpose, conjugate, Hermitian transpose, and $l_2$ norm of $\mathbf{a}$, respectively. For a scalar $B$, $\left | B \right |$ represents the $l_2$ norm of $B$. Furthermore, the expectation value is denoted as $\mathbb{E}[\cdot]$, and the determinant value is denoted as $\mathrm{det}(\cdot)$. Finally, $\mathcal{CN}(\mu,\Omega)$ denotes the complex Gaussian distribution with mean $\mu$ and variance $\Omega$.

\subsection{Paper Organization}
The rest of the paper is organized as follows. In Sec.\ref{sec:s2}, the communication and radar channel models are given and the proposed signaling architecture is reviewed. In Sec.\ref{sec:s3}, the decision bounds and regions are presented with a given example; then the derivation of the semi-analytical \gls{ber} expression is given. In Sec.\ref{sec:s4}, we derive closed-form expressions for the upper bound of the \gls{ber} of both \gls{zf} and \gls{jml} receiver. Furthermore, in Sec.\ref{sec:s5}, we consider the reliability of the communication system by analyzing the probability of system outage. Finally, the simulation results and discussion are presented in Sec.\ref{sec:s6}, and Sec.\ref{sec:s7} concludes the paper. The glossary of this paper is provided in the Table \ref{tab:1}.

\begin{table}[!t]
    \normalsize
    \centering
    \caption{List of Abbreviations}
    \label{tab:1}
    \printglossary[type=\acronymtype, style = mystyle, title = Acronyms]
\end{table}

\section{System model}
\label{sec:s2}

\begin{figure}[!t]
\centering
\includegraphics[scale=0.2]{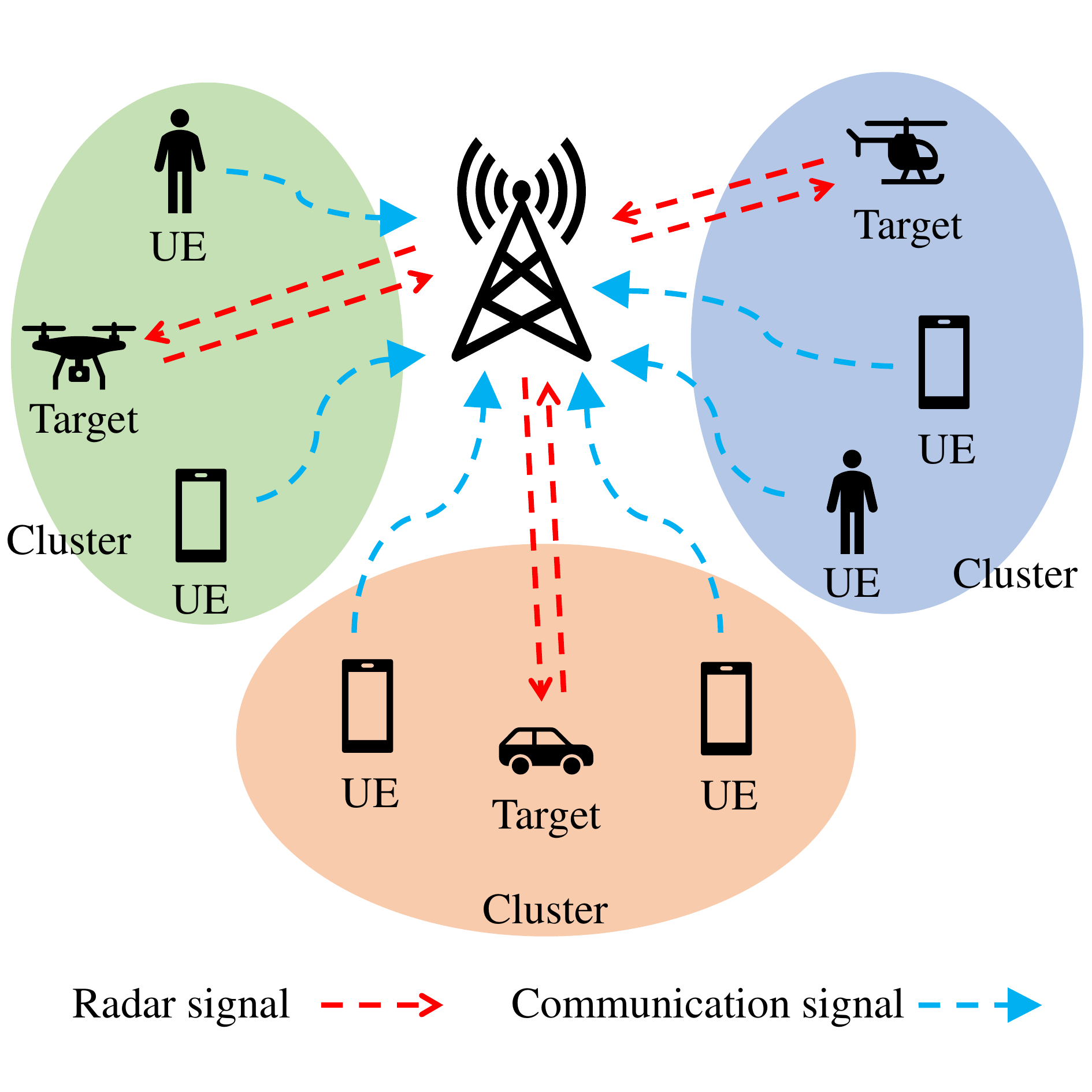}
\caption{System model.}
\label{fig:2-1}
\end{figure}

This work considers a system in which a \gls{noma}-based \gls{bs} simultaneously operates uplink data communication and radar sensing. For simplicity and without loss of generality, we assume that each cluster accommodates $K=2$ \glspl{ue} and a single radar target, as presented in Fig.~\ref{fig:2-1}. In addition, each \gls{ue} is equipped with a single antenna. The \gls{bs} is designed with two distinct sets of antennas, one set is dedicated to communication and the other for radar operations. Furthermore, the \gls{bs} has $M$ antennas to receive both communication and radar echoes.

\subsection{Communication Channel Model}
In this study, the communication channel between the $k$th \gls{ue} and the $m$th \gls{bs} antenna is denoted as $h_{m,k} = \acute{h}_{m,k} \beta_k$, where $\acute{h}_{m,k} \sim \mathcal{CN}(0, 1)$ is the small-scale fading and $\beta_k$ is the large-scale fading factor of the $k$th \gls{ue}, which is given by $\beta _k = \mathit{d}_k^{- \alpha}$, where $\mathit d_{k}$ represents the distance between the $k$th \gls{ue} and \gls{bs}, and $\alpha$ represents the path loss exponent of the communication channel. Consequently,
\begin{equation}
\mathbf h_k = \left [ h_{1,k}, h_{2,k}, \dots, h_{M,k} \right ] ^{\dag} \in \mathbb{C}^{M \times 1}.
\label{eq:2-1}
\end{equation}
We assume that \gls{bs} has perfect knowledge of the \gls{csi}, the communication channel is time-invariant between channel estimates, and \gls{ue}$_1$ is closer to \gls{bs}, that is, $d_1 < d_2$ and $\mathbb{E}\left[\left \| \mathbf{h}_1 \right \|^2\right ] \geq \mathbb{E}\left[\left \| \mathbf{h}_2 \right \|^2\right]$. 

\subsection{Radar Channel Model}
For the case where one target is assigned for each resource block, the gain of the two-way radar channel can be calculated based on \cite[Eq. 2.8]{radar_gain}, 
\begin{equation}
\acute{g}=\frac{\lambda}{8R^{2}}\sqrt{\frac{G_{\mathrm{t}}G_{\mathrm{r}}\sigma}{\pi^{3}}}\label{eq:2-2}
\end{equation}
where $G_{\mathrm{t}}$ and $G_{\mathrm{r}}$ represent the gains of transmit and receive antennas, respectively, $\sigma$ is the \gls{rcs} of the target, $R$ is the distance from the target to the \gls{bs}, and $\lambda$ is the signal wavelength. Using the general expression of the radar channel given in \cite[Eq. 1]{radar_channel}, the radar channel between the target and the $m$th \gls{bs} antenna can be represented as
\begin{equation}
g_m = \acute{g} \mathrm{e}^{-j2 \pi \Theta} \mathrm{e}^{j2 \pi f_\mathrm{d} \mu T}
\label{eq:2-3}
\end{equation}
where $\mathrm{e}^{-j2 \pi \Theta}$ is the phase shift component, which consists of the distance information of the target, $\Theta = 2 R/c_0$, where $c_0$ is the speed of light. The second phase shift term $\mathrm{e}^{j2 \pi f_\mathrm{d} \mu t_0}$ contains the velocity information of the target, $f_\mathrm{d}$ is the Doppler shift, $f_\mathrm{d} = 2 v f_\mathrm{c} / c_0$, where $v$ is the velocity of the target, and $f_\mathrm{c}$ is the signal carrier frequency. Furthermore, $T$ is the duration of the symbol, and $\mu$ is the symbol index. Therefore, the radar channel vector is given by
\begin{equation}
\mathbf g = \left [ g_1, g_2, \dots, g_M \right ]^{\dag} \in \mathbb{C}^{M \times 1}.
\label{eq:2-4}
\end{equation}

\subsection{The Signaling Model Under Consideration}

\begin{figure}[!t]
\centering
\includegraphics[width=0.9\linewidth]{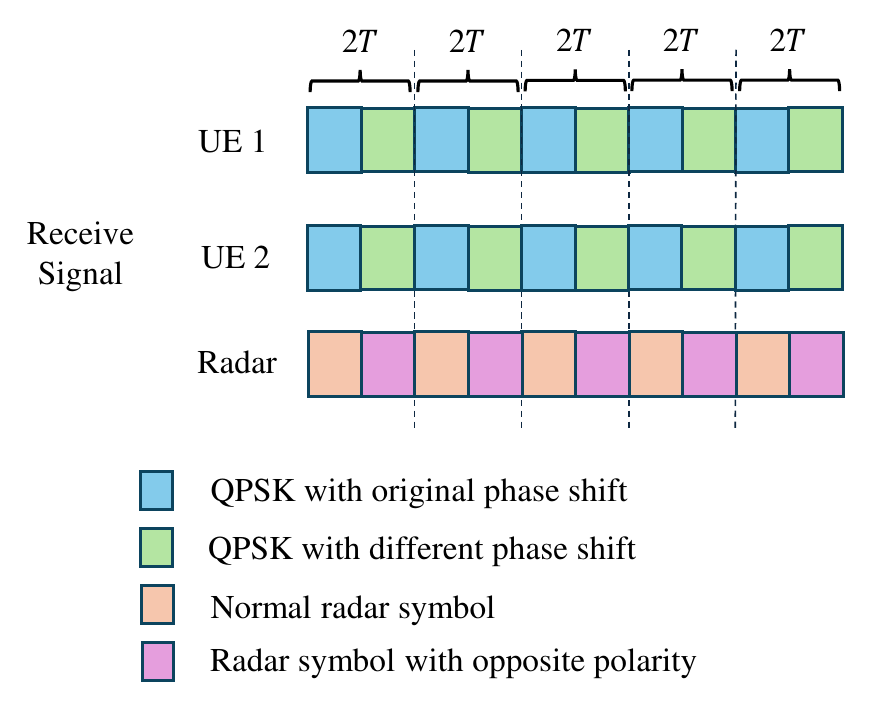}
\caption{The proposed signalling model of the \gls{isac} system.}
\label{fig:2-2}
\end{figure}

Conventional \gls{noma} detectors operate under the strict constraint that the power difference between the signals received from different users must exceed a specific threshold \cite{iraqi-PA}. However, in practical scenarios, such a condition is not always feasible, given the fluctuating nature of the signal power received. To address this challenge, a novel \gls{isac} signaling model is proposed in \cite{my}, with the assumption that the radar and \gls{ue} channels remain constant for the duration of two consecutive radar symbols \cite{coherence_time}. As shown in Fig.~\ref{fig:2-2}, to ensure auto-cancelation of the radar signal at the receiving end, the radar transmitter emits each symbol twice, but alternating its polarity in the second symbol period, e.g. $s_{\mathrm{r}}$ in the first period and $-s_{\mathrm{r}}$ in the second. Therefore, the signal received at the $m$th \gls{bs} antenna over two $T$ can be defined as \cite{MR-NOMA}, 
\begin{equation}
\hspace{-2mm}y_{m}=\left\{ 
\begin{array}{cc}
\sum\limits_{k=1}^{2}\sqrt{P_{k}}h_{m,k}x_{k,1}+\sqrt{P_{\mathrm{r}}}%
g_{m}s_{\mathrm{r}}+n_{m,1}, ~ t\in T_{1} \\ 
\sum\limits_{k=1}^{2}\sqrt{P_{k}}h_{m,k}x_{k,2}-\sqrt{P_{\mathrm{r}}}%
g_{m}s_{\mathrm{r}}+n_{m,2}, ~ t\in T_{2}%
\end{array}%
\right. 
\label{eq:2-5}
\end{equation}
\noindent where $n_{m,t}$ represents the noise in the $t$th period of the $m$th receive antenna, which follows the Gaussian distribution with zero mean and $\sigma_n^2$ variance. Furthermore, $P_k$ denotes the transmit power of the $k$th \gls{ue}, the total communication power is $P_{\mathrm{com}} = P_1 + P_2$, and $P_{\mathrm{r}}$ denotes the transmit power of the radar signal. In addition, $x_{k,t}$ denotes the transmitted communication symbol of the $k$th \gls{ue} at the $t$th period. By adding the received signals of both periods, we can obtain
\begin{align}
y_m = \sqrt{P_1} h_{m,1} x'_1 + \sqrt{P_2} h_{m,2} x'_2 + n'_{m},
\label{eq:2-6}
\end{align}
\noindent where $x'_1 = x_{1,1} + x_{1,2}$, $x'_2 = x_{2,1} + x_{2,2}$ and $n'_m = n_{m,1} + n_{m,2}$. Therefore, the radar signal is canceled regardless of its power level, and the remaining signals are pure communication signals over two symbol periods. After combing the signals of all $M$ antennas, the signal received over $2T$ is given by

\begin{align}
\mathbf y_{2T} & = 
\begin{bmatrix}
h_{1,1}\!\! & h_{2,1}\!\! & \cdots \!\! & h_{M,1} \\ 
h_{1,2}\!\! & h_{2,2}\!\! & \cdots \!\! & h_{M,2}
\end{bmatrix}^{\dag}
\begin{bmatrix}
\sqrt{P_1}x'_{1} \\ 
\sqrt{P_2}x'_{2}
\end{bmatrix} 
+ 
\begin{bmatrix}
n'_{1}\!  \\ 
\vdots\!  \\ 
n'_{M}
\end{bmatrix} \nonumber
\\
& = \sqrt{P_k} \mathbf H \mathbf x + \mathbf n.
\label{eq:2-7}
\end{align}

\begin{figure}[!t]
\centering
\subfigure[$\theta_\mathrm{r} = \pi/2$]{
    \begin{minipage}[t]{0.23\textwidth}
        \centering
        \includegraphics[width=1\textwidth]{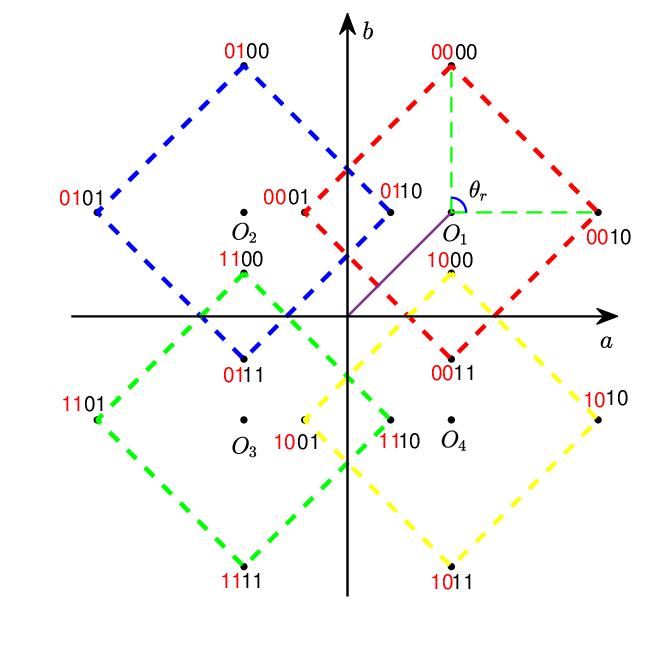}
    \end{minipage}}
\subfigure[$\theta_\mathrm{r} = \pi/3$]{
    \begin{minipage}[t]{0.23\textwidth}
        \centering
        \includegraphics[width=1\textwidth]{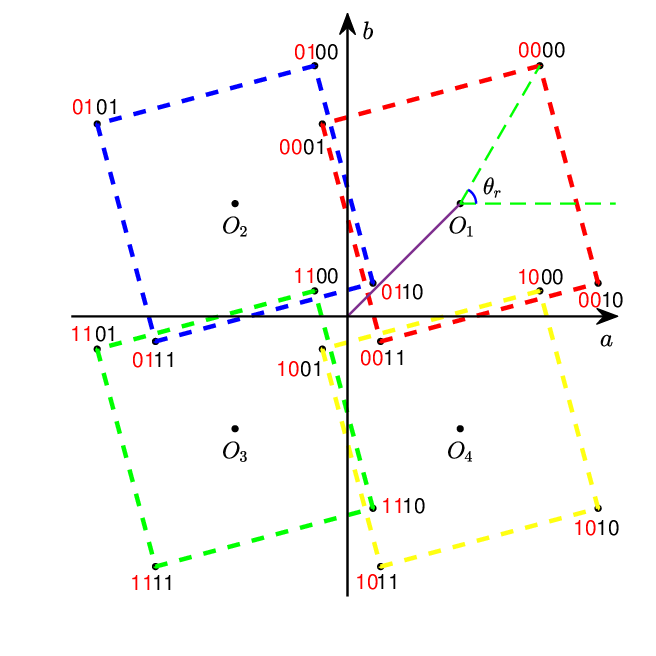}
        \end{minipage}}
\caption{The constellation diagram of symbols combined 2$T$ with different $\theta_\mathrm{r}$.}
\label{fig:2-3}
\end{figure}

It should be noted that, since we combine symbols over two $T$ for both \glspl{ue}, a detection ambiguity will occur if two symbols are of opposite polarity, such as $x_{1,1}=-x_{1,2}$, resulting in cancelation of the communication signal at the receiver. To resolve the ambiguity problem, we propose using alternate phase shifts on adjacent communication symbols. Fig.~\ref{fig:2-3} illustrates the different phase shifts with \gls{qpsk} modulation as an example, where $\theta_\mathrm{o}$ represents the original phase shift angel, and $\theta_\mathrm{r}$ represents the rotated phase shift angel. Specifically, the communication symbols in $T_1$ are selected from the original phase shift modulation, and the communication symbols in $T_2$ are selected from a constellation with the different phase shifts. By setting $\theta_\mathrm{r}$ to a fixed value, the \glspl{ue}' data are guaranteed to remain non-zero over two $T$ periods. For this to work, the \gls{bs} must have prior knowledge of the transmission protocol.

Compared to conventional \gls{noma}, the proposed \gls{isac} system does not require much increase in complexity, as it only requires three additional operations, which are to: sum over two $T$, invert the radar symbol every other $T$, and multiply each transmitted symbol by $\mathrm{e}^{\left(j(\theta_\mathrm{r} - \theta_\mathrm{o})\right)}$ every other $T$.

\section{ BER analysis of the ZF receiver}
\label{sec:s3}

As the decoding order in the \gls{isac} systems is difficult to determine, \gls{sic} receivers may not be applicable. On the one hand, if the system first decodes the communication signal, the communication \gls{asr} will be limited by interference from the sensing signal. On the other hand, it is not always possible to first decode and remove the radar signal, since the power of the received radar signal may be lower than that of the communication signal due to the two-way transmission \cite{SIC}.

By introducing a new matrix ${\mathbf H} = \left [\mathbf h_1, \mathbf h_2 \right ]$, the \gls{zf} receiver is denoted by $\mathbf W_{\mathrm{ZF}} \in \mathbb{C}^{2 \times M}$, which is given by
\begin{equation}
\mathbf W_{\mathrm{ZF}} = \left (\mathbf H^H \mathbf H \right )^{-1} \mathbf H^H,
\label{eq:3-1-1}
\end{equation}
\noindent or the pseudo-inverse of the channel matrix $\mathbf H$. The signal after the \gls{zf} receiver is given by
\begin{align}
\hspace{-3mm}\hat{\mathbf y}_{\mathrm{ZF}} = \mathbf W_{\mathrm{ZF}} \mathbf{y}_{2T} 
&= \mathbf W_{\mathrm{ZF}} \sum_{k = 1}^{K} \sqrt{P_k} \mathbf h_k x'_k + \mathbf W_{\mathrm{ZF}} \mathbf{n}_\mathrm{2T} \nonumber \\ 
&= \begin{bmatrix}
\sqrt{P_1} x'_1 + \mathbf w_{1} \mathbf{n}_\mathrm{2T}\\ 
\sqrt{P_2} x'_2 + \mathbf w_{2} \mathbf{n}_\mathrm{2T}
\end{bmatrix},
\label{eq:3-1-2}
\end{align}
\noindent since $\left \| \mathbf w_{k} \mathbf h_k \right \|^2 = 1$ and $\left \| \mathbf w_{k} \mathbf h_j \right \|^2 = 0$. After the \gls{zf} receiver, the \gls{ml} can be used to decode the signal of each \gls{ue}. By dividing the received signals to the power of each \gls{ue}, the signal sent to the \gls{ml} processor is given by
\begin{equation}
\hat{\mathbf y} =
\begin{bmatrix}
x'_1 + \mathbf w_{1} \mathbf{n}_\mathrm{2T} / \sqrt{P_1} \\ 
x'_2 + \mathbf w_{2} \mathbf{n}_\mathrm{2T} / \sqrt{P_2}
\end{bmatrix},
\label{eq:3-1-3}
\end{equation}
\noindent where the only disturbance encountered during each \gls{ue}'s detection process is due to noise. The power of noise of each \gls{ue} is given by 
\begin{equation}
P_{N}^{(1)} =  \frac{2 \sigma_n^2 \left [ \left (\mathbf H^H \mathbf H \right )^{-1} \right ]_{1,1}} { P_1 },
\label{eq:3-1-4}
\end{equation}

\begin{equation}
P_{N}^{(2)} = \frac{2 \sigma_n^2 \left [ \left (\mathbf H^H \mathbf H \right )^{-1} \right ]_{2,2}} { P_2 },
\label{eq:3-1-5}
\end{equation}
where $\mathbb{E} \left [\left | \mathbf{n}_\mathrm{2T} \right |^2 \right ] = 2 \sigma_n^2$ because the noise power is computed over $2T$.

\begin{figure}[!t]
\centering
\subfigure[Annotation]{
    \begin{minipage}[t]{0.23\textwidth}
        \centering
        \includegraphics[width=1\textwidth]{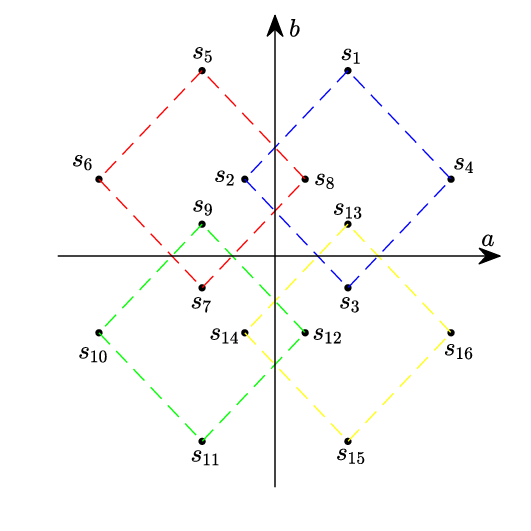}
    \end{minipage}}
\subfigure[Decision bounds]{
    \begin{minipage}[t]{0.23\textwidth}
        \centering
        \includegraphics[width=1\textwidth]{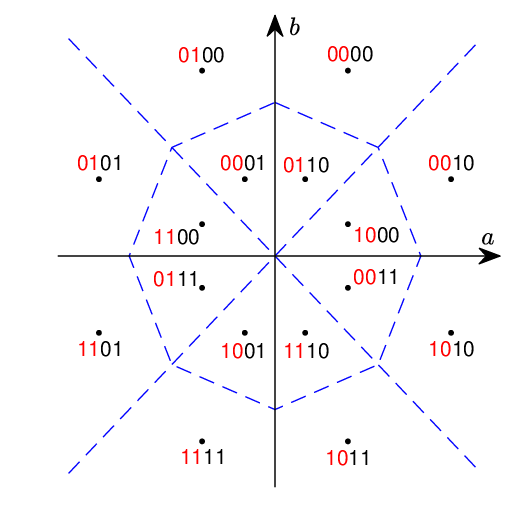}
        \end{minipage}}
\caption{A possible symbol annotation and decision bounds}
\label{fig:3-1}
\end{figure}

To simplify the discussion, the possible annotation for the symbols combined over $2T$ of each \gls{ue} are given in in Fig.~\ref{fig:3-1}(a). Let vector $\mathbf s$ denotes all $16$ possible combined symbols, $a_n$ and $b_n$ are the x coordinate and y coordinate of $n$th combined symbol, respectively. Therefore, for $n = 1, 2, \dots, 16$, $\mathbf s_n = a_n + b_n i$, $a_n$ and $b_n$ are given by
\begin{equation}
a_n = \cos \left( \left\lceil \frac{n}{4} \right\rceil \frac{\pi}{2} - \frac{\pi} {4} \right) + \cos \left[ \frac{\left(2n-1\right) \pi}{4} + \theta_\mathrm{r} \right],
\label{eq:3-1-6}
\end{equation}

\begin{equation}
b_n = \sin \left( \left\lceil \frac{n}{4} \right\rceil \frac{\pi}{2} - \frac{\pi} {4} \right) + \sin \left[ \frac{\left(2n-1\right) \pi}{4} + \theta_\mathrm{r} \right], 
\label{eq:3-1-7}
\end{equation}
\noindent where $\left\lceil \cdot \right\rceil$ denotes the ceiling operation. Recall \eqref{eq:3-1-3}, by sending each element of $\hat{\mathbf y}$ to the \gls{ml} receiver, the symbols can be decoded by applying the decision bounds shown in Fig.~\ref{fig:3-1}(b). Hence, the \gls{ml} receiver for the $k$th \gls{ue} is given by

\begin{align}
\left \{ \hat{x}_{k,1}, \hat{x}_{k,2} \right \} &= \underset{x_{k,1}, x_{k,2} \in \mathbf s} {\arg\min} \left | \hat{y}_k - x_{k,1} - x_{k,2} \right|^2 \nonumber\\
&= \underset{s_n \in \mathbf s} {\arg\min} \left | \hat{y}_k - s_n \right|^2.
\end{align}
\label{eq:3-1-8}

Furthermore, the \gls{pdf} of $\hat{\mathbf y}_k$ given information $x_{k,1}, x_{k,2}$ is
\begin{equation}
f \left( \hat{y}_k | x_{k,1}, x_{k,2} \right ) = \frac{1}{\pi P_{N}^{(k)}} \mathrm{e}^{-\frac{\left| \hat{y}_k - x_{k,1} - x_{k,2} \right| ^{2}} {P_{N}^{(k)}}}.
\label{eq:3-1-9}
\end{equation}

\begin{figure}[!t]
\centering
\includegraphics[scale=0.7]{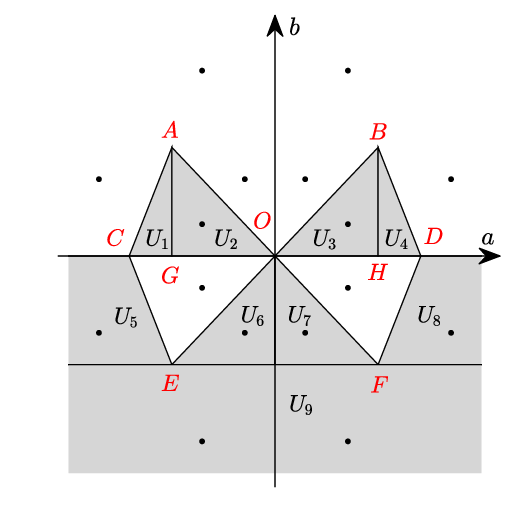}
\caption{The error decision regions.}
\label{fig:3-2}
\end{figure}

We now focusing on the \gls{ber} analysis. For instance, consider the symbol $s_8$ with the combined bit sequence `$0110$'. In this case, the desired \gls{msb} is `$0$', and an error detection event would occur if the \gls{msb} is mistakenly detected as `1'. The regions associated with this error detection correspond to areas $U_1$ to $U_9$, as shown in Fig.~\ref{fig:3-2}.

It is clear that the correct transmitted joint symbol set is $E_\mathrm{c} = \left [s_1, s_2, \dots, s_8 \right ]$, and the error decoded symbol set is $E_\mathrm{e} = \left [s_9, s_{10}, \dots, s_{16} \right ]$. Let $P \left (E_\mathrm{c} | s_n \right )$ be the probability that the \gls{msb} of the symbol is '0,' thus $P \left (E_\mathrm{c} | s_n \right ) = 1/8$. Recall \eqref{eq:3-1-6}, \eqref{eq:3-1-7} and \eqref{eq:3-1-9}, the \gls{pdf} of error detection is given by 
\begin{equation}
f \left( \hat{y}_k | s_n \right ) = \frac{1}{\pi P_{N}^{(1)}} \mathrm{e}^{-\frac{\left| \hat{y}_k - s_n \right| ^{2}} {P_{N}^{(1)}}} = f_{n(a)}(a-a_n) f_{n(b)}(b-b_n).
\label{eq:3-1-10}
\end{equation}
\noindent where $f_{n(a)}(a-a_n) = \frac{1} {\sqrt{\pi P_{N}^{(1)}}} \mathrm{e}^{-a^{2}/P_{N}^{(1)}}$ and $f_{n(b)}(b-b_n) = \frac{1} {\sqrt{\pi P_{N}^{(1)}}} \mathrm{e}^{-b^{2}/P_{N}^{(1)}}$ due to the isotropic symmetry of white Gaussian noise. Therefore, \gls{ber} for the $k$th \gls{ue} is given by
\begin{align}
P_{\mathrm{b},k} &= \sum_{n=1}^{8} \int_{U} P(E_\mathrm{c} | s_n) f \left( \hat{y}_k | s_n \right ) d \hat{\mathbf y}_k \nonumber \\
&= \frac{1}{8}\sum_{n=1}^{8} \int\int_{U} f_{n(a)}(a-a_{n}) f_{n(b)}(b-b_{n})dadb.
\label{eq:3-1-11}
\end{align}

Furthermore, the regions $U_1, U_2, \dots, U_9$ are given by
\begin{subequations}
\begin{align}
& U_1: \left \{ \left (a,b \right )|~ a_C \leq a \leq a_A, b_C \leq b \leq k_{AC}\left (a - a_C \right ) \right \} \\
& U_2: \left \{ \left (a,b \right )|~ a_G \leq a \leq a_O, b_O \leq b \leq k_{OA}\left (a - a_O \right ) \right \} \\
& U_3: \left \{ \left (a,b \right )|~ a_O \leq a \leq a_H, b_O \leq b \leq k_{OB}\left(a - a_O \right) \right \} \\
& U_4: \left \{ \left (a,b \right )|~ a_H \leq a \leq a_D, b_D \leq b \leq k_{BD}\!\left (a - \! a_D \right) \right \} \\
& U_5: \left \{ \left (a,b \right )|~ a \leq \left (b - b_E \right ) / k_{CE} + a_E, b_E \leq b \leq b_C \right \} \\
& U_6: \left \{ \left (a,b \right )|~ a_E \leq a \leq a_O, b_E \leq b \leq k_{OE}\left (a - a_O \right ) \right \} \\
& U_7: \left \{ \left (a,b \right )|~ a_O \leq a \leq a_F, b_F \leq b \leq k_{OF}\left (a - a_O \right ) \right \} \\
& U_8: \left \{ \left (a,b \right )|~ \!\left (b - b_F \right ) \!/ k_{DF} + a_F \leq a, b_F \leq b \leq b_D \right \} \\
& U_9: \left \{ \left (a,b \right )|~ b \leq b_F \right \},
\end{align}
\label{eq:3-1-12}
\end{subequations} 
where $a_A$ means the coordinate of $A$ on x-axis, $b_A$ means the coordinate of $A$ on y-axis, $k_{AC}$ means the slope of line $AC$. Due to the symmetry of the error regions and constellation diagram of $\mathbf s_n$, it is easy to get $(a_B, b_B) = (-a_A, b_A) = (a_F, -b_F) = (-a_E, -b_E)$, $a_B = a_H = -a_G$, $a_D = -a_C$, $k_{AC} = k_{DF} = -k_{BD} = -k_{CE}$, and $k_{OB} = k_{OE} = -k_{OA} = - k_{OF} = 1$.  By using the Q function $Q(x) = \frac{1}{\sqrt{2\pi}} \int_{x}^{\infty} \exp \left(-\frac{u^{2}}{2}\right) du$, then substituting \eqref{eq:3-1-12} into \eqref{eq:3-1-11} and simplifying them, the \gls{ber} for \gls{ue}$_1$ is given in \eqref{eq:3-1-13} where $\rho_k=\sqrt{P_{N}^{(k)}/2}$.

\begin{figure*}[tb!]
\begin{multline}
P_{\mathrm{b},k}=\frac{1}{8}\sum_{n=1}^{8}\left\{
1+Q\left( \frac{-b_{n}}{\rho _{k}}\right) \left[ Q\left( \frac{a_{D}+a_{n}}{%
\rho _{k}}\right) -Q\left( \frac{a_{D}-a_{n}}{\rho _{k}}\right) \right]
+Q\left( \frac{b_{B}+b_{n}}{-\rho _{k}}\right) \left[ Q\left( \frac{%
a_{B}+a_{n}}{-\rho _{k}}\right) -Q\left( \frac{a_{B}-a_{n}}{\rho _{k}}%
\right) \right] \right\}  \\
-\frac{1}{8\sqrt{\pi P_{N}^{(k)}}}\Bigg \{\int_{-b_{B}}^{0}\left[ Q\left( \frac{%
b+b_{B}}{-\rho _{k}k_{AC}}-\frac{a_{B}+a_{n}}{\rho _{k}}\right) +Q\left( 
\frac{b+b_{B}}{\rho _{k}k_{AC}}+\frac{a_{B}-a_{n}}{\rho _{k}}\right) \right] 
\mathrm{e}^{-\frac{(b-b_{n})^{2}}{P_{N}^{(k)}}}db \\
+\int_{-a_{D}}^{-a_{B}}Q\left( \frac{k_{AC}(a+a_{D})-b_{n}}{\rho _{k}}%
\right) \mathrm{e}^{-\frac{(a-a_{n})^{2}}{P_{N}^{(k)}}}da+\int_{-a_{B}}^{0}\left[
Q\left( \frac{a+b_{n}}{-\rho _{k}}\right) +Q\left( \frac{a-b_{n}}{\rho _{k}}%
\right) \right] \mathrm{e}^{-\frac{(a-a_{n})^{2}}{P_{N}^{(k)}}}da \\
+\int_{0}^{a_{B}}\left[ Q\left( \frac{a-b_{n}}{\rho _{k}}\right) +Q\left( 
\frac{a+b_{n}}{-\rho _{k}}\right) \right] \mathrm{e}^{-\frac{(a-a_{n})^{2}}{%
P_{N}^{(k)}}}da+\int_{a_{B}}^{a_{D}}Q\left( \frac{-k_{AC}(a-a_{D})-b_{n}}{\rho _{k}%
}\right) \mathrm{e}^{-\frac{(a-a_{n})^{2}}{P_{N}^{(k)}}}da\Bigg \}. 
\label{eq:3-1-13}
\end{multline}
\hrulefill
\vspace*{4pt}
\end{figure*}

The \gls{ber} expression is based on the random variable $\chi_k = \left[\left(\mathbf H^H \mathbf H\right)^{-1}\right]_{k,k}$. The $\chi_k $ follows the inverse-Gamma distribution \cite{Gamma_pdf}, which \gls{pdf} is given by 
\begin{equation}
f(\chi_k; \alpha_{\mathrm{ZF}}, \beta_{\mathrm{ZF},k}) = \frac{\beta_{\mathrm{ZF},k}^ {\alpha_{\mathrm{ZF}}}} {\Gamma (\alpha_{\mathrm{ZF}})} (1/\chi_k) ^ {\alpha_{\mathrm{ZF}} +1 } \exp(-\beta_{\mathrm{ZF},k} / \chi_k),
\label{eq:3-1-14}
\end{equation}
\noindent where $\alpha_{\mathrm{ZF}} = M-K+1$, $\beta_{\mathrm{ZF},k} = 1 / \beta_k$, and $\Gamma (\cdot)$ means the Gamma function. Therefore, after averaging over the inverse Gamma distribution, the semi-analytical \gls{ber} for the $k$th \gls{ue} is given by
\begin{equation}
P'_{\mathrm{b},k} = \int_0^{\infty} P_{\mathrm{b},k} ~ f\left(\chi_k; \alpha_{\mathrm{ZF}}, \beta_{\mathrm{ZF},k}\right) d\chi_k.
\label{eq:3-1-15}
\end{equation}

\section{Closed-from upper bound BER expression}
\label{sec:s4}

\subsection{Properties and rules of the outage probability evaluation}

With the fixed phase shift $\theta_\mathrm{r} = \pi/2$, we summarize the key properties when considering the outage probability as follows.

\begin{figure}[!t]
\centering
\subfigure[\gls{msb}]{
    \begin{minipage}[t]{0.23\textwidth}
        \centering
        \includegraphics[width=1\textwidth]{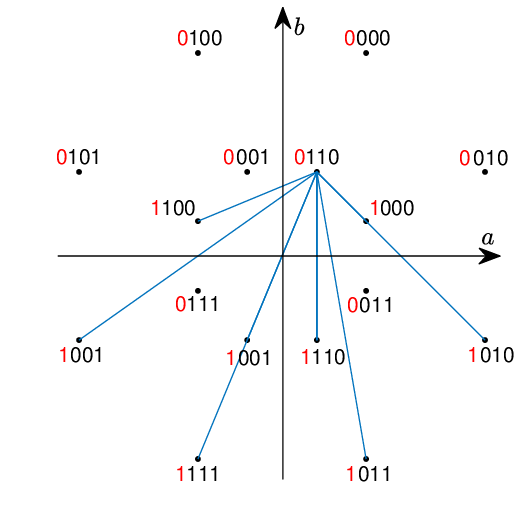}
    \end{minipage}}
\subfigure[\gls{lsb}]{
    \begin{minipage}[t]{0.23\textwidth}
        \centering
        \includegraphics[width=1\textwidth]{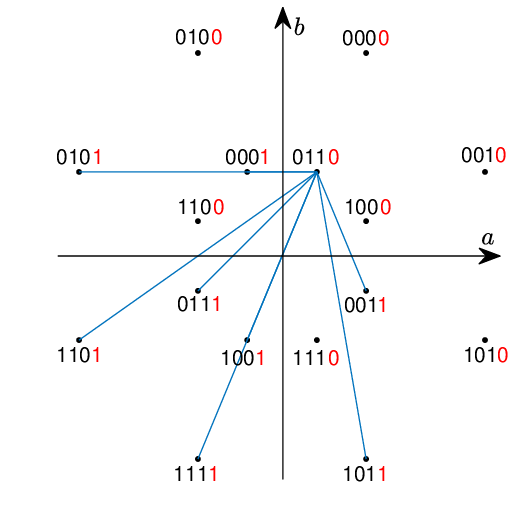}
        \end{minipage}}
\caption{The set for minimum Euclidean distances when correct value equal to 0. (a) \gls{msb}. (b) \gls{lsb}.}
\label{fig:4-1}
\end{figure}

\begin{enumerate}

\item \textit{Symmetry}. For each individual bit, only one case needs to be considered, that is, the correct bit equal to '0' or '1'. As shown in Fig.~\ref{fig:3-1}, the value for each single bit can only be '0' or '1'. Thus, when considering each independent bit, when the correct value is '0', the set of minimum Euclidean distances between the correct and wrong signals $\mathbf s_n$ is exactly the same as when the correct value is '1', since the new constellation diagram is symmetric about the coordinate axis.

\item \textit{Identically}. For each bit of any superimposed signal $\mathbf s_n$, the set of minimum Euclidean distances is the same, so only one bit need to be considered. As shown in Fig.~\ref{fig:4-1}, the error detection distances for the \gls{msb} and \gls{lsb} of the bit sequence '0110' are illustrated separately. It is easy to verify that the distance sets for the \gls{msb} and the \gls{lsb} are the same.

\end{enumerate}

Therefore, based on these two properties, for $k$th \gls{ue}, only one case for one bit needs to be considered. For simplicity, we consider the situation that the correct value for the \gls{msb} of the superimposed signal is '0'.

\subsection{The ZF receiver}

As shown in Fig.~\ref{fig:3-1}, the symbols with the \gls{msb} equal to '0' are $\mathbf s_1, \mathbf s_2, \cdots, \mathbf s_8$, and the rest of the symbols $\mathbf s_9, \mathbf s_{10}, \cdots, \mathbf s_{16}$ are the symbols with \gls{msb} equal to '1'. Let $d^{\mathrm{ZF}}_{c,w}$ represent the distance between the symbols $\mathbf s_c$ and $\mathbf s_w$. Wrong detection occurs when $n > d^{\mathrm{ZF}}_{c,w}/2$, that is, 
\begin{equation}
\begin{aligned}
p(n &> ~d^{\mathrm{ZF}}_{c,w}/2) \\ 
&= \int_{d^{\mathrm{ZF}}_{c,w}/2}^{\infty} \frac{1}{\sqrt{\pi P_{N}^{(k)}}} \mathrm{exp} \left [\frac{-v^2}{P_{N}^{(k)}} \right] dv 
= Q\left(\frac{d^{\mathrm{ZF}}_{c,w}}{\sqrt{2 P_{N}^{(k)}}}\right),
\end{aligned}
\label{eq:4-1-1}
\end{equation}
\noindent since $n$ follows zero mean and variance $P_{N}^{(k)} / 2$. Furthermore, the distance $d^{\mathrm{ZF}}_{c,w}$ when the \gls{msb} is equal to '0' is given by
\begin{equation}
d^{\mathrm{ZF}}_{c,w} = \left | \mathbf s_c - \mathbf s_w  \right |, c \in \{1,2,\dots,8\}, w \in \{9,10,\dots,16\}.
\label{eq:4-1-2}
\end{equation}

Therefore, the upper bound of \gls{ber} for the $k$th \gls{ue} with the \gls{msb} equal to '0' is given by 
\begin{equation}
P^{\mathrm{upper}}_{\mathrm{ZF},k}(\mathrm{first~bit = 0}) \leq \frac{1}{8} \sum_{c=1}^8 \sum_{w=9}^{16} Q\left(\frac{d^{\mathrm{ZF}}_{c,w}}{\sqrt{2 P_{N}^{(k)}}} \right ).
\label{eq:4-1-3}
\end{equation}

As we discussed previously, $P_{N}^{(k)}$ is based on the random variable $\chi_k = \left[\left(\mathbf H^H \mathbf H\right)^{-1}\right]_{k,k}$, which follows the inverse-Gamma distribution. The unconditional upper bound \gls{ber} for the $k$th \gls{ue} is given in \eqref{eq:4-1-4}, where $A = \sqrt{\frac{P_k ({d^{\mathrm{ZF}}_{c,w}})^2}{4 \sigma_n^2}}$, and ${}_2 F_1(\cdot)$ is the Hypergeometric function.

\begin{figure*}[!htb]
\begin{align}
P^{\mathrm{upper}}_{\mathrm{ZF},k} \!=\! \frac{1}{8} \sum_{c=1}^8 \sum_{w=1}^8 \int_0^{\infty} & \!Q\!\left(\frac{d^{\mathrm{ZF}}_{c,w}} {\sqrt{2 P_{N}^{(k)}}}\right) \!\! f\left(\chi_k; \alpha_{\mathrm{ZF}}, \beta_{\mathrm{ZF}}\right) \! d \chi_k \!=\! \frac{1}{8} \sum_{c=1}^8 \sum_{w=1}^8 \int_0^{\infty}\!\! Q\!\left(\frac{A}{\sqrt{\chi_k}}\right) \! \frac{\beta_{\mathrm{ZF}}^{\alpha_{\mathrm{ZF}}}} {\Gamma (\alpha_{\mathrm{ZF}})} \left(\frac{1}{\chi_k}\right) ^ {\alpha_{\mathrm{ZF}} +1 } \!\! \exp\left(\!-\frac{\beta_{\mathrm{ZF}}}{ \chi_k}\!\right)\! d \chi_k \nonumber \\
& = \frac{1}{8} \sum_{c=1}^8 \sum_{w=1}^8 \left [ 0.5 - \frac{\Gamma(0.5 + \alpha_{\mathrm{ZF}}) ~ {}_2 F_1\left(0.5, 0.5+\alpha_{\mathrm{ZF}}; 1.5; -A^2/\left(2 \beta_{\mathrm{ZF}}\right)\right)}{\sqrt{2 \pi \beta_{\mathrm{ZF}}} \Gamma \left(\alpha_{\mathrm{ZF}})\right /A} \right ]
\label{eq:4-1-4}
\end{align}
% \hrulefill
% \vspace*{4pt}
\end{figure*}

\subsection{The JML receiver}

The \gls{jml} receiver tends to decode the signal that has the highest probability which is given by
\begin{equation}
\hat{\mathbf x}' = \mathop{\arg\min}\limits_{x'_{1}, x'_{2} \in \mathbf s} \left \| \mathbf y_{2T} - \sqrt{P_1} {\mathbf h}_1 x'_{1} - \sqrt{P_2} {\mathbf h}_2 x'_{2} \right \|^2.
\label{eq:4-2-1}
\end{equation}

Since the \gls{jml} receiver decodes two \glspl{ue}' signals simultaneously, there are a total of $16 \times 16 = 256$ possibilities. The position of each possibility is given by $\sqrt{P_1}  \left \| \mathbf h_1 \right \| x'_1 + \sqrt{P_2} \left \| \mathbf h_2 \right \| x'_2 $. Similarly, assume that the correct \gls{msb} for \gls{ue}$_1$ is 0, the correct symbols for $x'_1$ are $s_1, s_2, \cdots, s_8$, and the wrong detection happens when $x'_1$ equals $s_9, s_{10}, \cdots, s_{16}$, regardless of the symbol $x'_2$. Therefore, by letting $s_c, s_p$ to be the correct signals, and $s_w, s_q$ to be the wrong signals for \gls{ue}$_1$ and \gls{ue}$_2$, respectively, the distances between the correct and wrong symbol positions for \gls{jml} receiver are given by
\begin{equation}
\begin{aligned}
&d^{\mathrm{JML},1}_{c,w} = \left \| \sqrt{P_1} \mathbf h_1 \left(s_c - s_w \right) + \sqrt{P_2} \mathbf h_2 \left(s_p - s_q\right) \right \|, \\
& c \in \{1,2,\dots,8\}, w \in \{9,10,\dots,16\}, p,q \in \{1,2,\dots,16\}.
\end{aligned}
\label{eq:4-2-2}
\end{equation}

Based on \eqref{eq:4-1-1} and \eqref{eq:4-2-2}, the upper bound for \gls{ue}$_1$ with the \gls{jml} receiver is given in \eqref{eq:4-2-3}, which is the conditional upper bound for given $\mathbf h_1$ and $\mathbf h_2$. With the help of the property of the Gaussian random vector, it is easy to get $\sqrt{P_1} \mathbf h_1 (s_c - s_w ) + \sqrt{P_2} \mathbf h_2 (s_p - s_q) \sim \mathcal{CN}\left(\mathbf{0}, \left ( P_1\beta_1 \left | s_c - s_w \right |^2 + P_2 \beta_2 \left | s_p - s_q \right |^2 \right) \cdot \mathbf{I}_M\right)$. Given the new variable, 

\begin{figure*}[!t]
\begin{equation}
P^{\mathrm{upper}}_{\mathrm{JML},1}(\mathrm{first~bit = 0}) 
= \frac{1}{8} \sum_{c=1}^{8} \sum_{w=9}^{16} \sum_{p=1}^{16} \sum_{q=1}^{16} Q \left (  \frac{\left \| \sqrt{P_1} \mathbf h_1 (s_c - s_w) + \sqrt{P_2} \mathbf h_2 (s_p - s_q) \right \|} {\sqrt{4 \sigma_n^2}} \right )
\label{eq:4-2-3}
\end{equation}
\hrulefill
\vspace*{4pt}
\end{figure*}

\begin{equation}
\delta^1_{c,w,p,q} = \frac{\left({d^{\mathrm{JML},1}_{c,w}}\right)^2} {4 \sigma_n^2},
\label{eq:4-2-4}
\end{equation}
\noindent then $\delta^1_{c,w,p,q}$ follows Erlang distribution with mean $M \eta^1_{c,w,p,q}$ and variance $M \left({\eta^1_{c,w,p,q}}\right)^2$ \cite{JML}, where $\eta^1_{c,w,p,q}$ is given by
\begin{equation}
\begin{aligned}
& \eta^1_{c,w,p,q} = \frac{P_1\beta_1 \left | s_c - s_w \right |^2 + P_2\beta_2 \left | s_p - s_q \right |^2} {4 \sigma_n^2}, \\
& c \in \{1,2,\dots,8\}, w \in \{9,10,\dots,16\}, p,q \in \{1,2,\dots,16\}.
\end{aligned}
\label{eq:4-2-5}
\end{equation}

The probability density function of $\delta^1_{m,n,p,q}$ is given by 
\begin{equation}
f_{\delta^1_{c,w,p,q}}(\delta) = \frac{\delta^{M-1}}{\left(M-1\right)! ~ \left({\eta^1_{c,w,p,q}}\right)^M} \mathrm{exp}\left(- \frac{\delta}{\eta^1_{c,w,p,q}}\right).
\label{eq:4-2-6}
\end{equation}

Therefore, based on \eqref{eq:4-2-3}-\eqref{eq:4-2-6}, the unconditional upper bound of \gls{ber} for \gls{ue}$_1$ is given by \eqref{eq:4-2-7}. 

Similarly, we assume the first correct bit for \gls{ue}$_2$ is 0, the distances between the correct and wrong symbol positions are given by
\begin{equation}
\begin{aligned}
&d^{\mathrm{JML},2}_{c,w} = \left \| \sqrt{P_2} \mathbf h_2 (s_c - s_w ) + \sqrt{P_1} \mathbf h_1 (s_p - s_q) \right \|, \\
& c \in \{1,2,\dots,8\}, w \in \{9,10,\dots,16\}, p,q \in \{1,2,\dots,16\}.
\end{aligned}
\label{eq:4-2-7}
\end{equation}

Given a new variable, 
\begin{equation}
\delta^2_{c,w,p,q} = \frac{\left(d^{\mathrm{JML},2}_{c,w}\right)^2} {4 \sigma_n^2},
\label{eq:4-2-8}
\end{equation}
\noindent which follows the Erlang distribution with mean $M \eta^2_{c,w,p,q}$ and variance $M \left({\eta^2_{c,w,p,q}}\right)^2$ \cite{JML}, where $\eta^2_{c,w,p,q}$ is given by
\begin{equation}
\begin{aligned}
&\eta^2_{c,w,p,q} = \frac{P_2\beta_2 \left | s_c - s_w \right |^2 + P_1\beta_1 \left | s_p - s_q \right |^2} {4 \sigma_n^2}, \\
& c \in \{1,2,\dots,8\}, w \in \{9,10,\dots,16\}, p,q \in \{1,2,\dots,16\}.
\end{aligned}
\label{eq:4-2-9}
\end{equation}

Therefore, the unconditional upper bound of the \gls{ber} for the $k$th \gls{ue} with the \gls{jml} receiver is given in \eqref{eq:4-2-10}. 

\begin{figure*}[!t]
\begin{equation}
P^{\mathrm{upper}}_{\mathrm{JML},k} = \frac{1}{8} \sum_{c=1}^{8} \sum_{w=9}^{16} \sum_{p=1}^{16} \sum_{q=1}^{16} \int_0^{\infty} Q\left(\sqrt{\delta}\right) f_{\delta^k_{c,w,p,q}}(\delta) ~ d \delta
= \frac{1}{8} \sum_{c=1}^{8} \sum_{w=9}^{16} \sum_{p=1}^{16} \sum_{q=1}^{16} \frac{ \Gamma(2M) {}_2 F_1\left(M, M+0.5; M+1; -\frac{2}{\eta^k_{c,w,p,q}}\right)}{2^M \left({\eta^k_{c,w,p,q}}\right)^{M} \Gamma(M)}.
\label{eq:4-2-10}
\end{equation}
\hrulefill
\vspace*{4pt}
\end{figure*}

\section{The outage probability}
\label{sec:s5}

\subsection{The ZF receiver}

The sum rate for $k$th \gls{ue} with \gls{zf} receiver over 2$T$ is given by
\begin{equation}
C^{\mathrm{ZF}}_k = \mathrm{log}_2 \left(1 + \frac{P_k  \left | x'_k \right |^2} {\left[\left(\mathbf H^H \mathbf H\right)^{-1}\right]_{k,k} 2 \sigma_n^2}\right) .
\label{eq:5-1}
\end{equation}
Therefore, the outage probability for $k$th \gls{ue} with the \gls{zf} receiver is given by
\begin{align}
P^{\mathrm{out}}_{\mathrm{ZF},k} &= 1 - \mathrm{Pr} \left ( \mathrm{log}_2 \left(1 + \frac{P_k  \left | x'_k \right |^2} {\left[\left(\mathbf H^H \mathbf H\right)^{-1}\right]_{k,k} 2 \sigma_n^2}\right) \geq C \right ) \nonumber \\
&= 1 - \mathrm{Pr} \left ( \left[\left(\mathbf H^H \mathbf H\right)^{-1}\right]_{k,k} \leq R_{\mathrm{ZF}} \right ),
\label{eq:5-2}
\end{align}
\noindent where $C$ is the minimum rate requirement for the $k$th \gls{ue}, and $R_{\mathrm{ZF}}$ is given by $\frac{P_k \left | x'_k \right |^2}{\left( 2^{C} - 1\right) 2 \sigma_n^2}$. As we discussed previously, $\chi_k = \left[\left(\mathbf H^H \mathbf H\right)^{-1}\right]_{k,k}$ follows the inverse-Gamma distribution, after simplification, the unconditional outage probability for the $k$th \gls{ue} with the \gls{zf} receiver is given by
\begin{equation}
P^{\mathrm{out}}_{\mathrm{ZF},k} \!=\! 1 -\! \int_{0}^{R_{\mathrm{ZF}}} \!\!\!\! f(\chi_k; \alpha_{\mathrm{ZF}}, \beta_{\mathrm{ZF},k}) ~d \chi_k \!=\! 1 - Q\left(\!\alpha_{\mathrm{ZF}},\frac{\beta_{\mathrm{ZF},k}}{R_{\mathrm{ZF}}}\! \right)\!,
\label{eq:5-3}
\end{equation}
\noindent where $Q\left(\alpha_{\mathrm{ZF}},\beta_{\mathrm{ZF},k}\right)$ is the regularized gamma function. Therefore, the outage probability of the \gls{zf} receiver can be calculated by averaging the sum of the outage probability of all the \glspl{ue}.

\subsection{The JML receiver}

The sum rate for the $k$th \gls{ue} with \gls{jml} receiver over 2$T$ is given by \cite{rate_ML}
\begin{equation}
C^{\mathrm{JML}}_k = -\mathrm{log} \left [ \mathrm{det} \left (  \left(1 + \frac{P_k \left | x'_k \right |^2}{2 \sigma_n^2} \mathbf{h}_k^H \mathbf{h}_k \right)^{-1} \right ) \right ].
\label{eq:5-4}
\end{equation}

Therefore, the outage probability for the $k$th \gls{ue} with the \gls{jml} receiver is given by
\begin{align}
&P^{\mathrm{out}}_{\mathrm{JML},k} \nonumber \\
&= \mathrm{Pr} \left ( -\mathrm{log} \left [ \mathrm{det} \left (  \left (1 + \frac{P_k \left | x'_k \right |^2}{2 \sigma_n^2} \mathbf{h}_k^H \mathbf{h}_k \right )^{-1} \right ) \right ] \leq C \right ) \nonumber \\
&= \mathrm{Pr} \left ( \mathbf h_k^H \mathbf h_k \leq R_{\mathrm{JML}} \right ),
\label{eq:5-5}
\end{align}
\noindent where $C$ is the minimum rate requirement for the $k$th \gls{ue}, and $R_{\mathrm{JML}}$ is given by $\frac{( 2^{C} - 1) 2 \sigma_n^2}{P_k \left | x'_k \right |^2}$. The random variable $\xi_k = \mathbf h_k^H \mathbf h_k$ follows the Gamma distribution, and the \gls{pdf} is given by
\begin{equation}
f(\xi_k; \alpha_{\mathrm{JML}}, \beta_{\mathrm{JML},k}) = \frac{\beta_{\mathrm{JML},k}^ {\alpha_{\mathrm{JML}}}} {\Gamma (\alpha_{\mathrm{JML}})} \xi_k ^ {\alpha_{\mathrm{JML}} - 1 } \exp(-\beta_{\mathrm{JML},k} \xi_k),
\label{eq:5-6}
\end{equation}
\noindent where $\alpha_{\mathrm{JML}} = M$ and $\beta_{\mathrm{JML},k} = 1/\beta_k$. After simplification, the unconditional outage probability for the $k$th \gls{ue} with the \gls{jml} receiver is given by
\begin{align}
P^{\mathrm{out}}_{\mathrm{JML},k} &=\int_{0}^{R_{\mathrm{JML}}} f(\xi_k; \alpha_{\mathrm{JML}}, \beta_{\mathrm{JML},k}) ~d \xi_k \nonumber \\
&= \frac{\gamma \left ( \alpha_{\mathrm{JML}}, \beta_{\mathrm{JML},k} R_{\mathrm{JML}} \right ) } {\Gamma(\alpha_{\mathrm{JML}})},
\label{eq:5-7}
\end{align}
\noindent where $\gamma(\cdot)$ represents the lower incomplete gamma function. Similarly, the outage probability of the \gls{jml} receiver can be calculated by averaging the sum of the \glspl{ue}' outage probabilities.

\section{Numerical results}
\label{sec:s6}

\begin{table}[!t]
\caption{System parameters.
\label{tab:2}}
\centering
\begin{tabular}{|c|c|c|} 
\hline
\textbf{Parameter}& \textbf{Value}& \textbf{Description} \\
\hline
$\alpha$    & 3.5                     & Path loss exponent \\
\hline
$f_c$       & 5.8\ GHz                & Carrier frequency \\
\hline
$B$         & 10\ MHz                 & Bandwidth \\
\hline
$T$         & 10\ $\mu$s              & Symbol duration \\
\hline
$p_{N0}$    & -174\ $\mathrm{dBm/Hz}$ & Noise spectral density \\
\hline
$M$         & 5                       & Number of \gls{bs} antennas \\
\hline
$G_{\mathrm{t}},G_{\mathrm{r}}$   & 2\ dB                   & Radar antenna gain \\
\hline
$\sigma$    & 0\ dBm$^2$              & \gls{rcs} \\
\hline
$\theta_o$  & $\pi$ / 4               & Original phase shift \\
\hline
$\theta_r$  & $\pi$ / 2               & Rotated phase shift \\
\hline
\end{tabular}
\end{table}

In our simulation, \gls{ue}$_1$ is randomly located in a 30 to 80 m communication range, and the distance of \gls{ue}$_2$ is set to $d_2 = 1.2 d_1$.  For the radar system, only one target is assigned to the same resource block which is randomly located between the radar sensing range of 10 and 50$m$ with speed 10$m/s$. In addition, \gls{qpsk} modulation is used for both communication and radar systems. All other parameters used during the simulation are shown in Table \ref{tab:2}.

\begin{figure}[!t]
\centering
\includegraphics[width=1\linewidth]{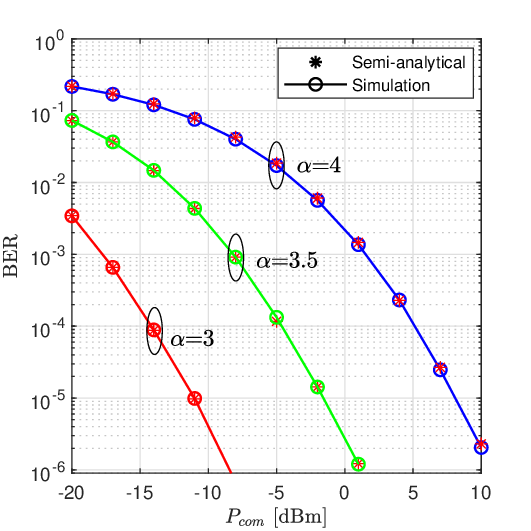}
\caption{The semi-analytical theoretical \gls{ber} performance as a function of \gls{ue} power $P_{\mathrm{com}}$ of \gls{zf} receiver and different path loss exponent $\alpha$.}
\label{fig:semi_ZF}
\end{figure}

In Fig.~\ref{fig:semi_ZF}, the performance of semi-analytical theoretical \gls{ber} analysis with the \gls{zf} receiver is presented. It is clear that the results of \eqref{eq:3-1-15} are perfectly matched with the simulation results. Additionally, when considering the different path loss exponents $\alpha$ of the channel, the theoretical results are still aligned with the simulations. Furthermore, increasing the value of $\alpha$ results in a worse channel for the \glspl{ue}. Since the \gls{zf} receiver is used, the noise will be amplified more when the channel is severe. Moreover, the use of the \gls{zf} receiver depends greatly on the quality of the channel. When the system has good channel conditions, there is a great performance difference in \gls{ber} at the same power $P_{\mathrm{com}}$.

\begin{figure}[!t]
\centering
\includegraphics[width=1\linewidth]{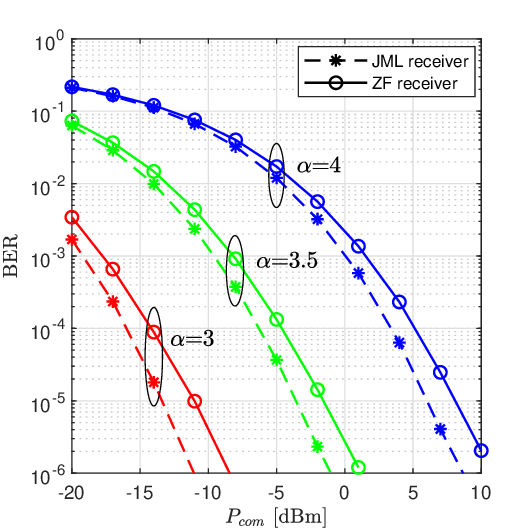}
\caption{The performance comparison of \gls{ber} between \gls{zf} and \gls{jml} receiver as a function of \gls{ue} power $P_{\mathrm{com}}$ with different path loss exponent $\alpha$.}
\label{fig:ZF_ML}
\end{figure}

Fig.~\ref{fig:ZF_ML} shows the performance difference between the \gls{zf} and \gls{jml} receivers. The \gls{ber} performances of the \gls{jml} receiver are consistently better than those of the \gls{zf} receiver. When the channel condition is relatively bad and in the lower power $P_{\mathrm{com}}$ region, the \gls{jml} receiver gives a very similar performance compared to the \gls{zf} receiver, since even though the \gls{jml} receiver does not amplify the power of the noise like the \gls{zf} receiver, the noise power is relatively high compared to the \glspl{ue}' power in this situation. When considering a system with good channel conditions, that is, when $\alpha = 3$, it can be found that the \gls{jml} receiver also performs better than the \gls{zf} receiver when the \glspl{ue} have lower power $P_{\mathrm{com}}$, mainly because the \gls{zf} receiver amplifies noise when eliminating inter-user interference. Furthermore, the \gls{jml} receivers generally have significant performance advantages over the \gls{zf} receivers, but in the proposed \gls{isac} system, this advantage is not obvious. The main reason is that in the proposed \gls{isac} system, the \gls{jml} receiver decodes the signals of two \glspl{ue} simultaneously, that is, it has $16 \times 16=256$ constellation points. In addition, since we use different phase shifts on adjacent symbol periods, the minimum Euclidean distance between symbols is not the same. Especially in the range close to the origin point, the distance between adjacent symbols is very small, thus reducing the performance of the \glspl{jml} receiver. For the \glspl{zf} receiver, it only needs to detect the data of a single \glspl{ue} separately, so each user only has 16 constellation points, and the minimum Euclidean distance between symbols is also larger than that of the \glspl{jml} receiver. So in the proposed \glspl{isac} system, the performance of the \glspl{jml} and \glspl{zf} receivers is relatively close.

\begin{figure}[!t]
\centering
\includegraphics[width=1\linewidth]{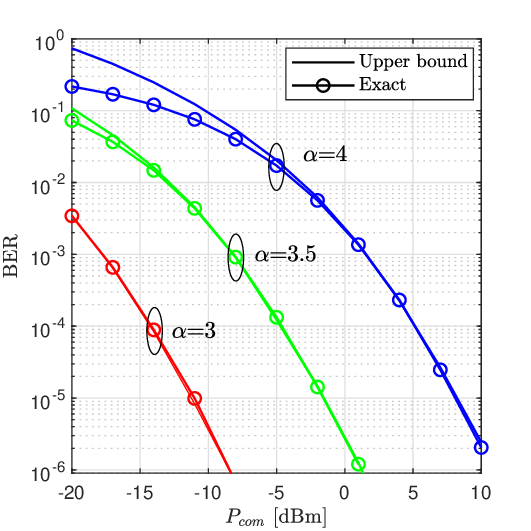}
\caption{The performance of \gls{ber} upper bound of \gls{zf} receiver as a function of \gls{ue} power $P_{\mathrm{com}}$ with different path loss exponent $\alpha$.}
\label{fig:upper_ZF}
\end{figure}

In Fig.~\ref{fig:upper_ZF}, the upper bounds of the \gls{ber} of the \gls{zf} receiver with different path loss exponents $\alpha$ are presented. We use the simulation results to demonstrate the exact performance of the \gls{ber}, and the upper bound of the \gls{ber} with the \gls{zf} receiver is given by \eqref{eq:4-1-4}. When \glspl{ue} have higher communication power $P_{\mathrm{com}}$, even if channel conditions are worse, for example $\alpha=4$, the upper bound of \gls{ber} is almost consistent with the exact \gls{ber} performance, giving a very close performance. When channel conditions improve, the performance of the upper bound of the \gls{ber} can be highly similar to that of the real \gls{ber} even at low power. Therefore, when the channel conditions are acceptable, the system can be evaluated by using the upper bound of \gls{ber}, thereby reducing the computational complexity and obtaining a very approximate system performance.

\begin{figure}[!t]
\centering
\includegraphics[width=1\linewidth]{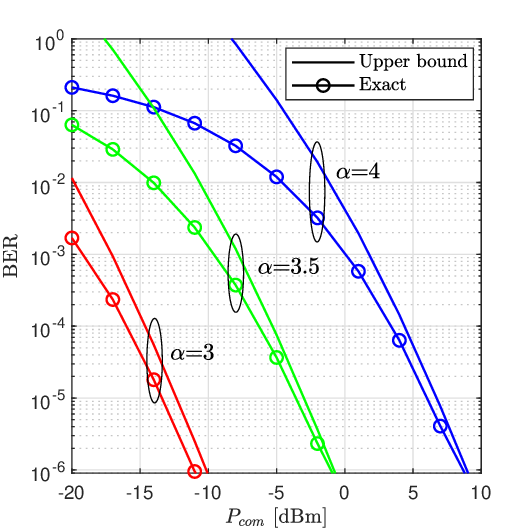}
\caption{The performance of \gls{ber} upper bound of \gls{jml} receiver as a function of \gls{ue} power $P_{\mathrm{com}}$ with different path loss exponent $\alpha$.}
\label{fig:upper_ML}
\end{figure}

In Fig.~\ref{fig:upper_ML}, the upper bounds of the \gls{ber} of the \gls{jml} receiver with different path loss exponents $\alpha$ are presented. Similarly, the simulation results are used to show the exact performance of the \gls{ber}, and the upper bound of the \gls{ber} with the \gls{jml} receiver is given by \eqref{eq:4-2-10}. As can be seen in the figure, under a variety of different channel conditions, the upper bound of the \gls{ber} of the \gls{jml} receiver does not fit well with the exact performance of \gls{ber} even in the region with higher communication power. Furthermore, bad channels will cause the difference between the upper bound of the \gls{ber} and the exact \gls{ber} to increase significantly in the lower communication power region. 

Furthermore, it should be noted that the complexity of calculating the upper bound of the \gls{ber} of the \gls{jml} receiver is much greater than that of the \gls{zf} receiver. It can be seen from \eqref{eq:4-1-4} and \eqref{eq:4-2-10} that since the \gls{jml} receiver decodes the signals of two \glspl{ue} in two symbol periods at the same time, the number of $d^{\mathrm{JML},k}_{c,w}$ to be considered is extremely large, while the \gls{zf} receiver eliminates the interference between \glspl{ue}, thus reducing the amount of $d^{\mathrm{ZF}}_{c,w}$ that needs to be considered, resulting in less computational complexity. Furthermore, since there are many possibilities for erroneously decoding signals when using the \gls{jml} receiver, this results in a large difference between the upper bound of the \gls{ber} of the \gls{jml} receiver and the exact \gls{ber}. When channel conditions worsen, it means a reduction in the \gls{snr} at the same power, once again increasing the difference between the upper bound and the exact \gls{ber}. In contrast, because the \gls{zf} receiver only considers the situation of each individual \gls{ue}, the possibility of error decoding is reduced, and the gap between the upper bound of the \gls{ber} and the exact value is smaller.

\begin{figure*}[!t]
\centering
\subfigure[]{
    \begin{minipage}[t]{0.32\textwidth}
        \centering
        \includegraphics[width=1\textwidth]{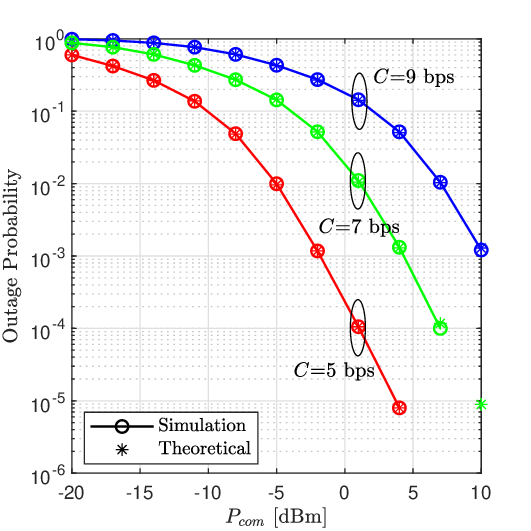}
    \end{minipage}}
\subfigure[]{
    \begin{minipage}[t]{0.32\textwidth}
        \centering
        \includegraphics[width=1\textwidth]{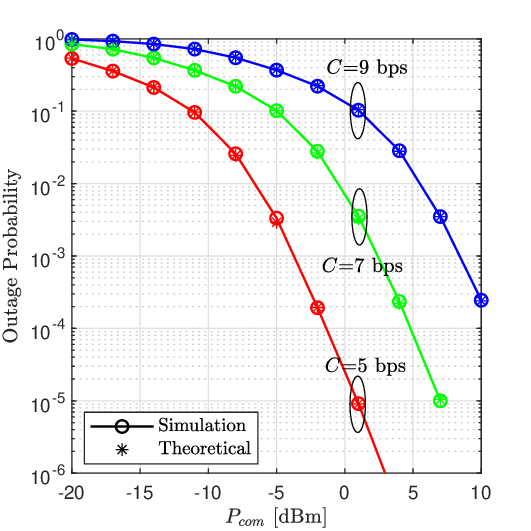}
        \end{minipage}}
\subfigure[]{
    \begin{minipage}[t]{0.32\textwidth}
        \centering
        \includegraphics[width=1\textwidth]{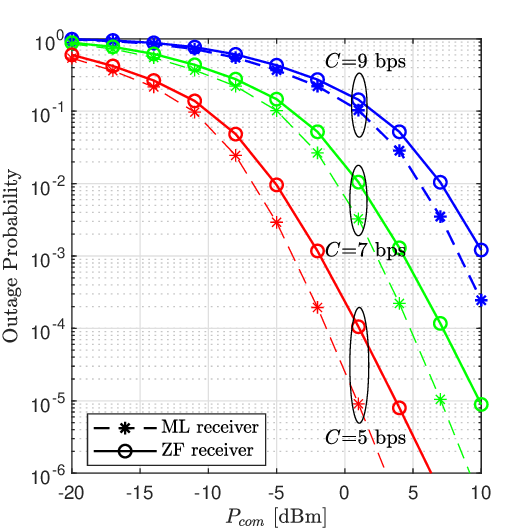}
        \end{minipage}}
\caption{The performance of outage probability as a function of communication power $P_{\mathrm{com}}$ with different minimum \gls{asr} requirement $C$. (a) \gls{zf} receiver. (b) \gls{jml} receiver. (c) Comparison between \gls{zf} and \gls{jml} receiver.}
\label{fig:out}
\end{figure*}

Fig.~\ref{fig:out}(a) presents the performance of the outage probability with the \gls{zf} receiver given the minimum \gls{asr} constraints $C = \left \{5,7,9 \right \}$bps. It shows that the analytical results given by \eqref{eq:5-3} are perfectly matched to the simulation results. Furthermore, it is clear that with an increase in the minimum rate requirement $C$ for the \glspl{ue}, the probability of the system being out of service increases for the same communication power $P_{\mathrm{com}}$. Similarly, Fig.~\ref{fig:out}(b) also demonstrates the performance of the outage probability with the \gls{jml} receiver given minimum \gls{asr} constraints $C = \left \{5,7,9 \right \}$bps. Again, the analytical results given in \eqref{eq:5-7} are in perfect agreement with the simulation results. Furthermore, we compare the performance of outage probability with the \gls{zf} and \gls{jml} receiver in Fig.~\ref{fig:out}(c). Clearly, the \gls{jml} receiver has better performance compared to the \gls{zf} receiver, and the difference between the \gls{zf} and the \gls{jml} receiver tends to be greater with increasing $P_{\mathrm{com}}$. The reason is that at the same value of $P_{\mathrm{com}}$, the communication \gls{asr} of the \gls{jml} receiver is also higher than that of the \gls{zf} receiver, as in \cite{my}. In addition, the \gls{zf} receiver will amplify the noise power when canceling the inter-user interference, leading to a lower \gls{sinr}; thus, the \gls{zf} receiver has a smaller communication \gls{asr} compared to the \gls{jml} receiver at the same communication power $P_{\mathrm{com}}$, which means a higher outage probability.

It should be noted that the computational complexity of the \gls{jml} receiver is $\mathcal{O}(Q^{2K})$, where $Q$ is the modulation order. Compared to the \gls{zf} receiver which has $\mathcal{O}(K^3)$ computational complexity due to matrix inversion operations, the \gls{zf} receiver obviously gives a lower system cost. Therefore, although the \gls{jml} receiver provides better performance in both \gls{ber} and outage probability, considering the computational cost and the difference in performance, it is better to choose the \gls{zf} receiver rather than the \gls{jml} receiver, which offers a balanced trade-off between system performance and computational complexity.

\section{Conclusions}
\label{sec:s7}

In this study, we explore the \gls{ber} and outage probability of a novel \gls{isac} of \gls{noma} based uplink \gls{iot} system, in which detection is done over two symbol periods. We develop a semi-analytical expression to calculate the exact \gls{ber} of the \gls{zf} receiver, and derive closed-form expressions for the upper \gls{ber} bounds of both the \gls{zf} and the \gls{jml} receivers. In addition, we derive closed-form expressions for the outage probability of both the \gls{zf} and \gls{jml} receivers. The results indicate a strong alignment between the semi-analytical \gls{ber} and closed-form outage probability expressions with the simulation results. We also examine the upper bound of the \gls{ber} under varying path loss exponents. Overall, the numerical results showed that the \gls{jml} receiver performs better in both the \gls{ber} and the outage probability, especially in higher \gls{snr} regions. However, the \gls{jml} receiver requires a much higher computational cost compared to the \gls{zf} receiver, and the performance of the \gls{jml} receiver is also degraded due to the use of different phase shifts over $2T$. Therefore, the \gls{zf} receiver is a balanced approach when considering the trade-off between the system performance and complexity.

\bibliographystyle{IEEEtran}
% \bibliography{IEEEabrv}


% \input{reference.bbl}
% \bibliographystyle{IEEEtran}
% \bibliography{IEEEabrv,reference}

\end{document}